\documentclass[twocolumn,superscriptaddress,amsmath,amssymb]{revtex4-2}
\usepackage{color}
\usepackage{graphicx}
\usepackage{subfigure}
\usepackage{booktabs}
\usepackage{dcolumn}
\usepackage{bm}
\usepackage{amsmath}
\usepackage{verbatim}
\usepackage{lineno}

\usepackage[colorlinks=true]{hyperref}
\hypersetup{
  colorlinks=true,
  linkcolor=blue,
  filecolor=magenta,
  citecolor=blue,
  urlcolor=blue,
}

\newcommand{\beq}{\begin{eqnarray}}
\newcommand{\eeq}{\end{eqnarray}}

\newcommand\YJ[1]{\textcolor{red}{ [YJ:\,#1]}}

\begin{document}

\title{A review on shear jamming}

\author{Deng Pan}
\affiliation{CAS Key Laboratory of Theoretical Physics, Institute of Theoretical Physics, Chinese Academy of Sciences, Beijing 100190, China}

\author{Yinqiao Wang}
    \affiliation{Research Center for Advanced Science and Technology, University of Tokyo, 4-6-1 Komaba, Meguro-ku, Tokyo 153-8505, Japan.}

\author{Hajime Yoshino}
\affiliation{Cybermedia Center, Osaka University, Toyonaka, Osaka 560-0043, Japan}
\affiliation{Graduate School of Science, Osaka University, Toyonaka, Osaka 560-0043, Japan}

\author{Jie Zhang}
    \affiliation{School of Physics and Astronomy, Shanghai Jiao Tong University, 800 Dong Chuan Road, 200240 Shanghai, China}
    \affiliation{Institute of Natural Sciences, Shanghai Jiao Tong University, 200240 Shanghai, China}
    
\author{Yuliang Jin}
    \email[Email address: ]{yuliangjin@mail.itp.ac.cn}
\affiliation{CAS Key Laboratory of Theoretical Physics, Institute of Theoretical Physics, Chinese Academy of Sciences, Beijing 100190, China}
\affiliation{School of Physical Sciences, University of Chinese Academy of Sciences, Beijing 100049, China}
\affiliation{Center for Theoretical Interdisciplinary Sciences, Wenzhou Institute, University of Chinese Academy of Sciences, Wenzhou, Zhejiang 325001, China}


\date{\today}

\begin{abstract}
Jamming is a ubiquitous phenomenon that appears in many soft matter systems, including granular materials, foams, colloidal suspensions, emulsions, polymers, and cells -- when jamming occurs, the system undergoes a transition from flow-like to solid-like states. Conventionally, the jamming transition occurs when the system reaches a threshold jamming density under isotropic compression, but recent studies reveal that jamming can also be  induced by shear. Shear jamming has attracted much interest in the context of non-equilibrium phase transitions, mechanics and rheology of amorphous materials. 
Here we review the phenomenology  of shear jamming and its related physics.  
We first describe basic observations obtained in experiments and simulations, and results from  theories.  Shear jamming is then demonstrated as a ``bridge" that connects the rheology of athermal soft spheres and thermal hard spheres. Based on a generalized jamming phase diagram, 
a universal description is provided for shear jamming in frictionless and frictional systems. We further review the isostaticity and criticality of the shear jamming transition, and the elasticity of shear jammed solids. The broader relevance of shear jamming is discussed, including its relation to other phenomena such as shear hardening, dilatancy, fragility, and discrete shear thickening.

\end{abstract}

\maketitle
\tableofcontents

\section{Introduction}

The jamming transition~\cite{liu1998nonlinear, makse2000packing, o2003jamming} is of particular interest in two research areas in physics.
(i) Non-equilibrium phase transitions. In a standard liquid-to-solid phase transition, i.e., crystallizaition,  the solid state is structurally ordered.
The jamming transition represents, perhaps the simplest,  paradigm of ``unusual" liquid-to-solid transitions, where  structural ordering appears to be absent -- another example of such transitions is the glass transition. 
It is the ``simplest", in the sense that  the interaction energy between constituting particles (e.g., granular particles) is generally many orders of magnitude greater than the thermal energy, making temperature effects negligible in many circumstances. As a result, the jamming transition behaves like a well defined athermal phase transition with unambiguous critical scalings -- this is in contrast to the case of glass transition, where the thermodynamic nature of the phase transition is generally hindered by complex  dynamics.   
Remarkably, recent progress has revealed deep connections between the jamming transition in athermal systems and the glass transition in supercooled liquids, following strategies (in particular, the mean-field replica theory) initially developed in the studies of spin glasses~\cite{parisi2010mean, charbonneau2011glass,
charbonneau2017glass, parisi2020theory}. 
It thus provides a unique opportunity to understand both transitions in a unified framework. 

 (ii) Physics of amorphous solids. The jammed states near the jamming transition belong to an ``unusual" type of solid, which are structurally disordered and marginally stable. 
Such properties correspond to anomalous behavior that breaks down several standard descriptions developed for crystalline solids. 
The marginal stability is associated with soft vibrational modes, in excess of the well-known Debye modes in crystals. These soft modes have extremely low  excitation energies beyond which the system loses stability, and have irregular spatial structures~\cite{liu2010jamming,  muller2015marginal}.
The response of a jammed state to mechanical perturbations, such as shear, is not purely affine as in crystals -- thus the seminal elasticity theory by Born and Huang~\cite{born1954dynamical} has to be extended with non-affine contributions included~\cite{maloney2004universal, wyart2005rigidity, lemaitre2006sum, karmakar2010athermal, zaccone2011network, yoshino2010emergence, yoshino2012replica, yoshino2014shear,  pan2022shear}. Because the system is  marginally stable, its elasticity is inevitably coupled to plasticity in the thermodynamic limit, making the linear shear modulus protocol-dependent~\cite{yoshino2014shear, nakayama2016protocol,  jin2017exploring}, and the fluctuations of non-linear shear moduli diverging~\cite{hentschel2011athermal, procaccia2016breakdown, biroli2016breakdown}.

Particles are commonly jammed by  increasing its volume fraction (density) with compression, during which the system remains isotropic. 
Isotropic jamming has been discussed in a number of reviews from various perspectives~\cite{
alexander1998amorphous, richard2005slow, makse2004statistical, van2009jamming,  liu2010jamming, parisi2010mean, torquato2010jammed,  wyart2012marginal,  bi2015statistical, tighe2010force, charbonneau2017glass, baule2018edwards, behringer2018physics}. 
Is it possible to jam  a system by anisotropic shear deformations  without increasing the density? 
If yes, does the above-mentioned physics change in shear jamming? 
Are there any connections 
between shear jamming and other rheological behavior of amorphous materials? 
 These are  the fundamental questions we aim to discuss in this review.

\begin{figure*}[!htbp]
  \centering
  \includegraphics[width=0.9\linewidth]{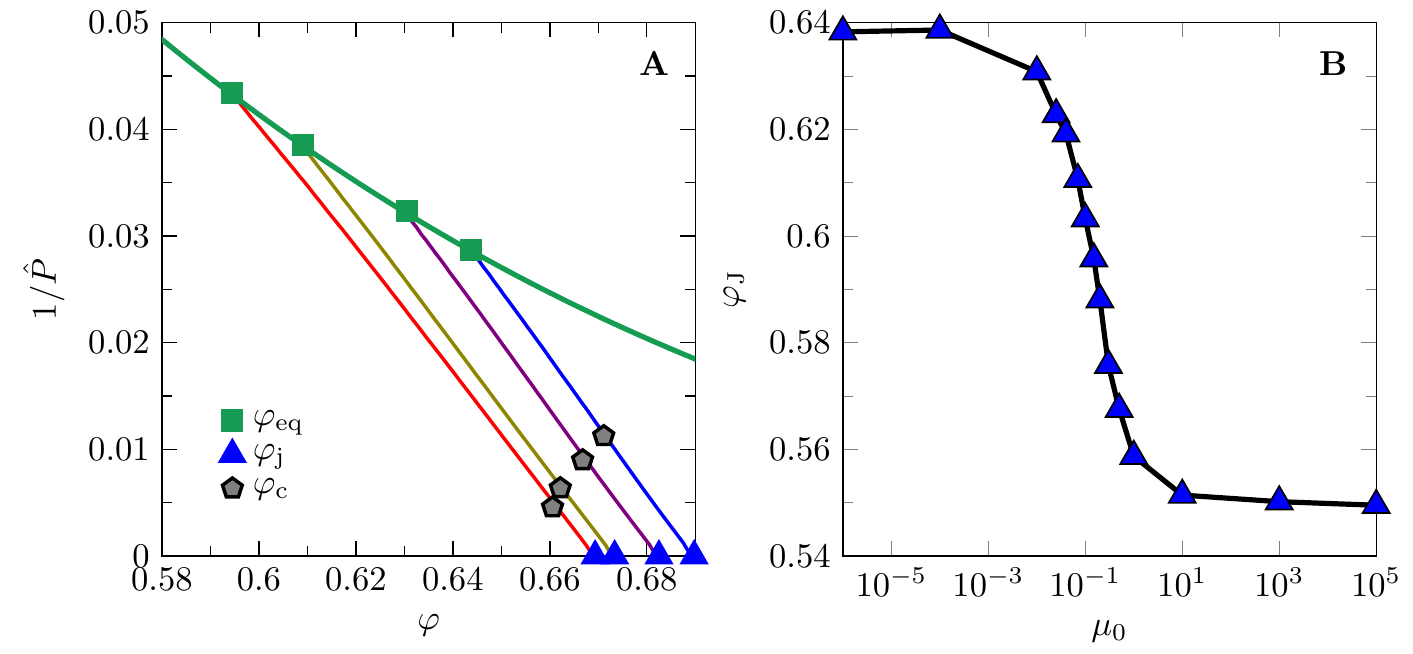}
  \caption{
  {\bf Non-unique isotropic jamming densities.}
  (A) Dependence of the isotropic jamming density $\varphi_{\rm j}$ on the parent liquid density $\varphi_{\rm eq}$ in simulated frictionless polydisperse thermal hard spheres (adapted from~\cite{jin2021jamming}). 
  The jammed states at $\varphi_{\rm j}$ are compression quenched from equilibrium liquid states at $\varphi_{\rm eq}$ that are prepared using the swap algorithm~\cite{berthier2016equilibrium}. The density $\varphi_{\rm c}$ separates shear yielding and shear jamming behavior in thermal hard sphere glasses (see Secs.~\ref{sec:frictionless_HS} and~\ref{sec:connection_athermal_thermal}).
  The green line represents the liquid equation of state. Other colored lines represent the glass equations of states, approximated by $\hat{P}(\varphi) \sim 1/(\varphi_{\rm j} - \varphi)$, where $\hat{P}$ is the reduced pressure of thermal hard spheres~\footnote{In thermal hard spheres, the pressure and stress are originated from particle collisions, or entropic effects; the value of the temperature $T$ is irrelevant since it just  fixes the unit of time. To avoid confusion, for thermal hard spheres we use the reduced pressure and stress, $\hat{P} = PV/Nk_{\rm B}T$ and $\hat{\sigma} = \sigma V/Nk_{\rm B}T$, where $V$, $N$ and $k_{\rm B}$ are the volume, number of particles and Boltzmann constant respectively. Note that in athermal ($T=0$) soft spheres, $\hat{P} = \hat{\sigma} = \infty$ by definition, and we always use the non-reduced  pressure and stress, $P$ and $\sigma$.}  
  (B) Dependence of  $\varphi_{\rm J}$  on the inter-particle friction coefficient $\mu_0$ in simulated monodisperse athermal soft spheres (adapted from ~\cite{song2008phase, wang2011jamming}).}
  \label{fig:J_line}
\end{figure*}

Shear jamming was 
 first observed in experiments of two-dimensional (2D) photoelastic disks over a decade ago~\cite{zhang2010jamming, bi2011jamming}: 
 jammed states were generated by applying a shear strain to initially stress-free configurations, at a  volume fraction $\varphi$ below the isotropic jamming-point (J-point) density $\varphi_{\rm J} \approx 0.84$~\cite{o2003jamming} of frictionless disks.   Thus shear deformations
 can facilitate the formation of many additional  mechanically stable, anisotropically jammed states  inaccessible by isotropic compression.
 Since then, this phenomenon has received much attention 
 in experiments~\cite{Waitukaitis-2012impact, 
 ren2013reynolds, zheng2014shear, 
 peters2016direct, zhao2019shear, zhao2022ultrastable}, simulations~\cite{vinutha2016disentangling, chen2018stress, bertrand2016protocol, baity2017emergent, 
 seto2019shear, kumar2016memory, das2020unified, babu2021dilatancy,
 kawasaki2020shear, jin2018stability, singh2020shear, jin2021jamming, giusteri2021shear, pan2022shear, peshkov2023comparison} and theories~\cite{urbani2017shear,  altieri2019mean, sarkar2013origin, sarkar2015shear, sarkar2016shear}, due to its rich physics and 
 close connections with geohazards and rheological applications such as discrete shear thickening~\cite{Brown-Jaeger-Review}.

 This review is organized as follows. In  Sec.~\ref{sec:J_plane}, we discuss 
 how  the phase space of jammed states is extended from isotropic to shear jamming. 
We then describe experimental observations in 
dry granular matter  and dense suspensions (Sec.~\ref{sec:experiments}), 
theoretical results (Sec.~\ref{sec:theory}) and numerical investigations by computer simulations (Sec.~\ref{sec:simulation}). 
In sec.~\ref{sec:connection_athermal_thermal}, we show the connection between the rheology of 
athermal soft spheres and thermal hard spheres via shear jamming.
A universal description of shear jamming is presented in Sec.~\ref{sec:phase_diagram} based on a generalized jamming phase diagram, for athermal systems with and without  friction. 
In Sec.~\ref{sec:scaling}, the critical scalings near the shear jamming transition are reviewed, and compared to those in isotropic jamming. In Sec.~\ref{sec:elasticity}, we focus on the elasticity of shear jammed solids.
In Sec.~\ref{sec:related},  we discuss 
rheological phenomena related to shear jamming, including shear dilatancy, shear hardening, fragile states, and discrete shear thickening. 
 An outlook is provided in Sec.~\ref{sec:outlook} with open questions.



 \section{From isotropic  to shear jamming: the jamming-point, jamming-line and jamming-plane}
 \label{sec:J_plane}

In this section, we review how the ensemble of jammed states has been successively extended. The first extension, from a {\it jamming-point} (J-point) to a {\it jamming-line} (J-line), is realized by taking into account the protocol-dependence (e.g., compression rate-dependence) of the isotropic jamming density. The J-line is then further extended into a {\it jamming-plane} (J-plane) by introducing anisotropy to the jammed configurations using shear protocols.

{(i) Jamming-point.} In three dimensions (3D), the densest packing of equal-sized spheres is achieved by  the face-centered cubic  or  the hexagonal close packing  structures, with a packing fraction  of  $\varphi_{\rm FCC} = \varphi_{\rm HCP} \simeq 0.74$~\cite{weaire2008pursuit, hales2017formal}. The ``random version'' of the sphere packing problem in 3D, however, remains  unsolved rigorously.  
Bernal et al. 
noticed that when ball bearings are poured inside containers, the packing fraction generally never exceeds the {\it random close packing} (RCP) density 
$\varphi_{\rm RCP} \approx 0.64$~\cite{bernal1960packing, scott1960packing}, even after shaking or kneading. 
It was then suggested that the RCP
corresponds to a new type of non-equilibrium phase transition, the jamming transition, occurring at a  jamming density $\varphi_{\rm J} \simeq \varphi_{\rm RCP} \approx 0.64$~\cite{liu1998nonlinear, makse2000packing, o2003jamming, song2008phase}. 


{(ii) Jamming-line.} Many studies have revealed that the jamming density is in fact non-unique -- 
isotropic jamming can occur over a continuous range of densities. 
In the absence of inter-particle friction, the jamming density depends on the protocol
~\cite{speedy1996distribution, torquato2000random, chaudhuri2010jamming, ozawa2012jamming}: a rapid compression procedure (compression rate $\dot{\varphi} \to \infty$) results in a minimum jamming density $\varphi_{\rm j}(\dot{\varphi}  \to \infty) =   \varphi_{\rm J} $~\cite{o2003jamming}~\footnote{We use $\varphi_{\rm j}$ for the non-uniquejamming density that depends on preparation protocols, and $\varphi_{\rm J}$ for the minimum jamming (J-point) density. When the inter-particle friction $\mu_0$ is non-zero, $\varphi_{\rm J}(\mu_0)$ is the minimum  jamming density for the given $\mu_0$.}, while in the infinitely slow annealing limit, the so-called  
{\it glass close packing} (GCP) density $\varphi_{\rm j}(\dot{\varphi} \to 0) = \varphi_{\rm GCP}$ is achieved, whose exact value remains unknown except for the asymptotic behavior in large dimensions~\cite{parisi2010mean, parisi2020theory}. 
Note that to define GCP, crystallization is assumed to be completely avoided. 
The dependence of $\varphi_{\rm j}$ on $\dot{\varphi}$ shall be interpreted in a general sense as the { protocol-dependence} of the jamming density $\varphi_{\rm j}$. For example,  athermal mechanical training can also generate various $\varphi_{\rm j}$ depending on protocol parameters~\cite{kumar2016memory, das2020unified,  kawasaki2020shear, babu2021dilatancy, pan2022nonlinear}. Another example is  the dependence of  $\varphi_{\rm j}$  on the density $\varphi_{\rm eq}$  of the parent equilibrium liquid state in  the {\it state-following} setup~\cite{franz1995recipes, rainone2016following} (see Fig.~\ref{fig:J_line}A);
in particular, GCP follows from the  {\it Kauzmann point} (or the {\it ideal glass transition}) at $\varphi_{\rm K}$~\cite{parisi2010mean, parisi2020theory}, which is proposed to be  the maximum density of liquid states~\cite{kauzmann1948nature}.

The isotropic jamming density also depends on the inter-particle (microscopic) friction coefficient $\mu_0$.
It is suggested that 
the RCP is reached in the zero-friction limit, $\varphi_{\rm J}(\mu_0 \to 0) =\varphi_{\rm RCP} $, and 
the {\it random loose packing} (RLP) in the infinite-friction limit, $\varphi_{\rm J}(\mu_0 \to 
\infty) = \varphi_{\rm RLP} \approx 0.55$~\cite{song2008phase, onoda1990random} (see Fig.~\ref{fig:J_line}B for the dependence of $\varphi_{\rm J}$ on  $\mu_0$). 
In the presence of inter-particle friction, the jamming density can still be increased by mechanical training such as athermal cyclic shear from the lowest jamming density $\varphi_{\rm J}(\mu_0)$  to a higher jamming density $\varphi_{\rm j} > \varphi_{\rm J}(\mu_0)$~\cite{pan2022nonlinear}.

\begin{figure}[!htbp]
  \centering
 \includegraphics[width=\linewidth]{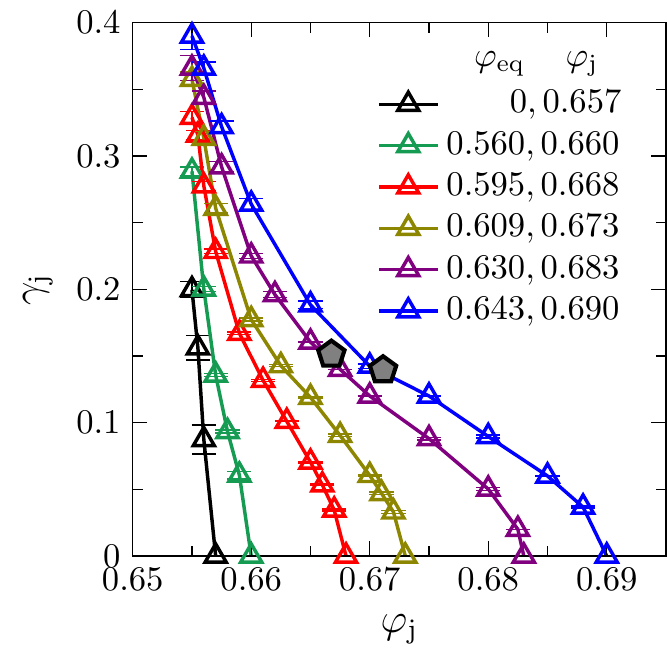}
  \caption{
  {\bf Jamming plane.}
  The J-plane is formed by a large
 number of shear jamming lines.
  A few examples of shear jamming lines are provided for the frictionless polydisperse sphere model~\cite{jin2021jamming}, with $\varphi_{\rm eq}$ and $\varphi_{\rm j}$ indicated in the legend.
  Each shear jamming line starts from an isotropic jamming point  at $(\varphi_{\rm j}, \gamma_{\rm j}=0)$ on the J-line, and asymptotically (in the thermodynamic limit) extends to  $(\varphi_{\rm J}, \gamma_{\rm j} \to \infty)$ upon shear. The isotropic states at $\varphi_{\rm j}$ are quenched from parent liquid states at $\varphi_{\rm eq}$ (see Fig.~\ref{fig:J_line}A).
  Along a shear jamming line, the point at $\varphi_{\rm c}$ (gray pentagon) separates shear yielding and shear jamming in thermal hard spheres~\cite{urbani2017shear}; the same point separates  reversible jamming and irreversible jamming~\cite{jin2021jamming}.
  The data are adapted from \cite{jin2021jamming}.
  }
  \label{fig:Jplane}
\end{figure}

{(iii) Jamming-plane.} When shear deformations are considered, the phase space of jammed states is further extended into a J-plane that spans density ($\varphi_{\rm j}$) and shear strain ($\gamma_{\rm j}$) axes~\cite{jin2021jamming}.
A shear jammed state can be obtained generally in the following way: starting from an isotropically jammed state at $\varphi_{\rm j}$, one first unjams the system by decompressing it to a density $\varphi$ in the range  $\varphi_{\rm J} < \varphi < \varphi_{\rm j}$, and then increases the shear strain deformation until the system jams again at a jamming strain $\gamma_{\rm j}$. In this way, any state point on the J-line can be extended into a line of shear-jammed states ({\it shear jamming line}) with different degrees of anisotropy, and the J-line can be systematically extended into a J-plane (see Fig.~\ref{fig:Jplane}). 


\begin{figure*}[!htbp]
  \centering
 \includegraphics[width=\linewidth]{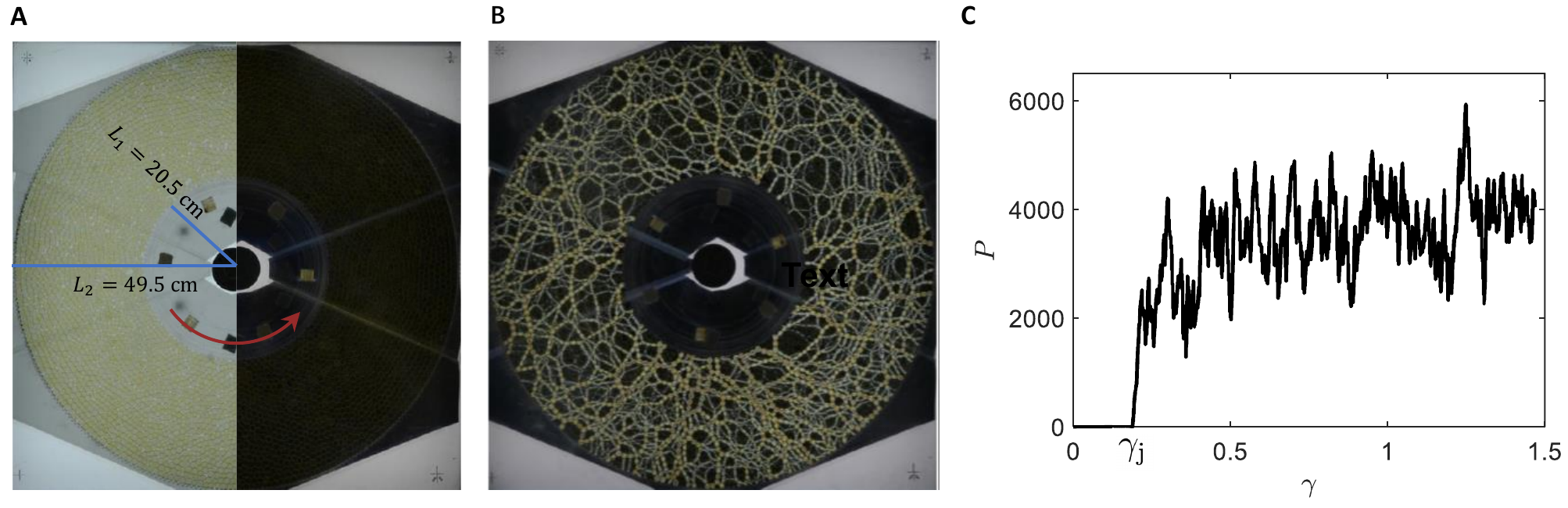}
  \caption{{\bf Shear jamming in granular experiments.}
  (A) Couette system used in the experiment of photoelastic disks, where the state is unjammed. (B) Force network of a shear-jammed state visualized by  image processing techniques. (C) Pressure-strain curve. Shear jamming, accompanied  with an abrupt onset of a finite pressure, can be identified at $\gamma_{\rm j} \approx 0.2$.
  }
  \label{fig:experiment}
\end{figure*}

\section{Experiments}
\label{sec:experiments}

 \subsection{Dry granular matter}
\label{sec:granular_matter}
An example of shear jamming in a two-dimensional  Couette system is given in Fig.~\ref{fig:experiment}. The Couette system consists of an annular cell between an inner circle 
and an outer concentric ring 
placed on a flat horizontal acrylic base plate. A horizontal layer of bidisperse photoelastic disks is randomly placed within the annular cell. 
The outer ring is fixed, whereas the inner circle can rotate freely, driven by a step motor. Typically, the outer ring and inner circle circumferences are designed to be gear-like, with regular teeth of the particle-size scale, which can enhance particle-wall driving at the inner wall and prevent particles from slipping at the outer wall. A circular polarizer is attached below the horizontal acrylic base plate to visualize the stress inside the disk layer. Meanwhile, 
another matched circular polarizer is placed at the top of the Couette cell below a high-resolution camera mounted 
above the system, which allows one to take stress images of force chains.

Once shear is applied to the layer of disks by rotating the inner wheel anti-clockwise, particles nearest to the wall are entrained to move, which  successively  entrain next thin layers of neighboring particles to move. In this way, the shearing motion of particles inside the annular is created. 
The rotation speed of the inner wheel is plodding, 
to mimic quasi-static shear deformations. 
Figure~\ref{fig:experiment}(A) shows stress-free  images for the unstrained (unjammed) initial state at a volume fraction $\varphi=0.818$, which is fixed during shear.
Beyond the jamming strain $\gamma_{\rm j} \approx 0.2$,
the system is shear-jammed with force chains percolating through the entire system, as shown in Fig.~\ref{fig:experiment}(B). 
Here, the strain $\gamma$ is defined as the ratio between the rotation distance of the inner wall and its radius.


To see how a non-zero pressure (stress) develops and changes with the strain, 
the pressure-strain curve is plotted in Fig.~\ref{fig:experiment}(C), where the pressure $P$ is quantified by the mean gradient square 
through calibration~\cite{howell1999stress}. 
Below $\gamma_{\rm j} \approx 0.2$, 
the system is unjammed with the pressure staying at zero; beyond $\gamma_{\rm j}$ the system develops a non-zero pressure increasing monotonically until a {\it steady state} is reached. 
Similar behavior can be observed in the stress-strain curve. 
Shear jamming is also accompanied by a jump of the coordination number $Z$ (average number of contacts per particle) from zero to a finite value satisfying the isostatic condition~\cite{babu2021dilatancy} (see Sec.~\ref{sec:isostaticity} for a discussion on the isostatic condition in frictionless systems). Thus the system becomes rigid after shear jamming, supported by a system-spanning force network as visualized in Fig.~\ref{fig:experiment}B.

\subsection{Dense suspensions}
\label{sec:colloids}

A suspension is a homogeneous mixture of solid particles (Brownian or non-Brownian) in a liquid.
Peters et al. performed a careful experiment of density-matched cornstarch suspension using a Couette system excluding the boundary-introduced compression effect, and observed that a fully jammed state can be reached purely by shear~\cite{peters2016direct}. 
In the experiment,  a sharp front between fluid-like and solid-like states develops and  propagates radially outwards, once a sudden rotation of the inner cylinder is applied to the suspension. After the front reaches the outer wall, the system fully jams while the inner cylinder continuous slipping. 
In the shear-jammed state, the suspension no longer behaves like a fluid: if a steel ball is dropped onto the suspension, it bounces back and stays on the surface   as long as shear is continuously applied.

The shear jamming behavior is seemingly  similar to the famous {\it discrete shear thickening} phenomenon, i.e., a sudden jump of the viscosity $\eta = \tau/\dot{\gamma}$ in its value  for several orders of magnitude when 
continuously increasing the strain rate $\dot{\gamma}$~\cite{Brown-Jaeger-Review, Morris-Review, ness2022physics}.
However,  the mechanical response of the shear-jammed state is dramatically different from that of a discrete shear thickening state; in the latter, a dropped ball sinks gradually into the suspension. In other words, the shear-jammed state behaves like a solid, while the  discrete shear thickening state behaves like a fluid. 
In Sec.~\ref{sec:fragile} , we discuss the connection between discrete shear thickening and  the {\it fragile states} that appear before the onset of shear jamming.

\section{Theories}
\label{sec:theory}
Shear jamming was studied theoretically in the frictionless thermal hard sphere model by Urbani and Zamponi, using mean-field replica calculations with the state-following formalism~\cite{urbani2017shear}. It is found that, depending on the density $\varphi$, a hard sphere glass can either shear yield or shear jam under the constant-volume condition. 
The separation occurs at a critical density $\varphi_{\rm c}$: for $\varphi_{\rm c} < \varphi < \varphi_{\rm j}$, shear jamming occurs, which is substituted  by shear yielding for  $\varphi < \varphi_{\rm c}$. This finding later leads to a {\it stability map} that unifies rheology and stability of hard sphere glasses~\cite{jin2018stability, altieri2019mean}.

The theoretical solution in Ref.~\cite{urbani2017shear} is exact in large dimensions (i.e., in the mean-field limit), but its relevance to finite-dimensional systems needs to be examined. Simulation  results of thermal hard spheres in three dimensions are  qualitatively consistent with these theoretical predictions~\cite{jin2017exploring, jin2018stability, jin2021jamming}, as discussed in Sec.~\ref{sec:frictionless_HS}. 
However, in simulations of athermal soft spheres, shear jamming is observed even for $\varphi < \varphi_{\rm c}$ (see Sec.~\ref{sec:connection_athermal_thermal}). This kind of shear jamming (called {irreversible shear jamming}~\cite{jin2021jamming}) is missing in the above theory. 


For athermal frictional systems, Sarkar  et al. have developed a theoretical framework based on a stress-only viewpoint~\cite{sarkar2013origin, sarkar2015shear, sarkar2016shear}. The emergence of rigidity at shear jamming is related to broken translational symmetry in force space, due to the  constraints of force and torque balance, the positivity of  normal forces, and the Coulomb criterion.

\begin{figure*}[!htbp]
  \centering
 \includegraphics[width = \linewidth]{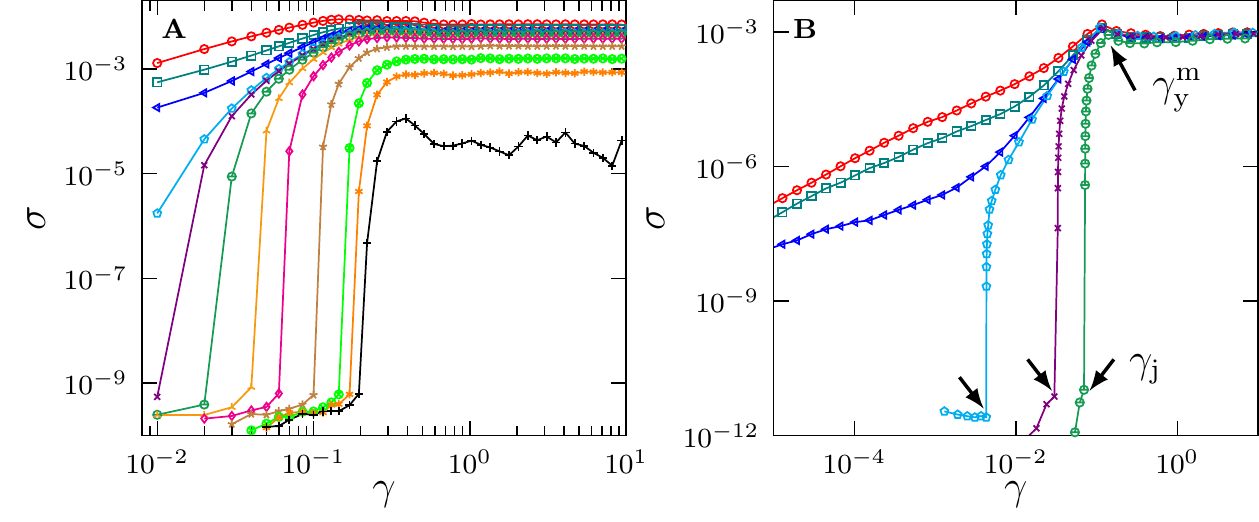}
  \caption{{\bf Shear jamming in simulations of 
  athermal soft spheres.} (A)
   Stress-strain curves for bidisperse frictional disks with $\mu_0 = 1.0$. Each curve is for a constant density: from left to right, $\Delta \varphi = \varphi - \varphi_{\rm j} = 2.6 \times 10^{-2}, 9.9 \times 10^{-3}, 3.9 \times 10^{-3}, 4.7 \times 10^{-4}, -2.0 \times 10^{-4}, -2.0 \times 10^{-3}, -5.0 \times 10^{-3}, -1.0 \times 10^{-2}, -2.0 \times 10^{-2}, -3.0 \times 10^{-2}, -3.5 \times 10^{-2}, -4.0 \times 10^{-2}$. The unstrained initial states are prepared by cyclic shear, with  $\varphi_{\rm j} = 0.835$.
  (B) Stress-strain curves (from left to right, $\Delta \varphi  = 4.1 \times 10^{-3}, 1.1 \times 10^{-3}, 4.5 \times 10^{-5}, -5.0 \times 10^{-5}, -2.0 \times 10^{-3}, -8.0 \times 10^{-3}$) for polydisperse frictionless spheres. The unstrained initial states are prepared by thermal annealing using the swap algorithm, with  { $\varphi_{\rm eq} = 0.643$} and $\varphi_{\rm j} = 0.69$.
  }
  \label{fig:frictional_simulation}
\end{figure*}

\section{Simulations}
\label{sec:simulation}

\subsection{Athermal soft spheres}
\label{sec:frictionless_SS}
\subsubsection{Frictional spheres}

The experimental observation of shear jamming in dry granular matter can be reproduced in computer simulations of athermal frictional particles. For a given  inter-particle friction coefficient $\mu_0$, the normal force $f_{\rm n}$  and the tangential force $f_{\rm t}$ obey the Coulomb's condition $f_{\rm t} \leq \mu_0 f_{\rm n}$. In general, $f_{\rm n}$ is computed from   a 
power-law potential, $V (\delta) \sim \delta ^\alpha$ where $\delta$ quantifies the linear overlap between two particles and $V=0$ when two particles are not in contact (in particular, $\alpha = 2$  and $5/2$ for the repulsive harmonic
and Hertzian interactions respectively). 
Before the equality in Coulomb's law is reached, i.e., before the microslip at the contact surface occurs, the incremental tangential force is proportional to the relative tangential displacement 
between two contacting spheres, 
as originally shown by Mindlin~\cite{mindlin1949campliance}. 
Athermal quasistatic shear is simulated by dissipating the energy after each incremental strain step until mechanical equilibrium is reached. The initial unstrained states are mechanically trained by cyclic shear.
Simulation results  are obtained for a frictional bidisperse model in two dimensions (Fig.~\ref{fig:frictional_simulation}A and Fig.~\ref{fig:schematic_PD}). 
More details on the model and simulation method can be found in Ref.~\cite{pan2022nonlinear}. Note that all simulation data presented in this review are averaged over many independent samples.

Similar to the experimental results (see Fig.~\ref{fig:experiment}C), for $\varphi < \varphi_{\rm j}$ the simulated stress-strain curves show the onset of shear jamming at $\gamma_{\rm j}$  (see Fig.~\ref{fig:frictional_simulation}A). Here $\gamma_{\rm j}$ depends on $\Delta \varphi = \varphi - \varphi_{\rm j}$; it increases monotonically with  $|\Delta \varphi|$ as shown in Fig.~\ref{fig:Jplane}. When $\varphi > \varphi_{\rm j}$, the system is isotropically jammed without shear; for such systems {\it shear hardening} can be observed~\cite{pan2022shear} (see Sec.~\ref{sec:other}).
In both cases ($\Delta \varphi<0$ and $\Delta \varphi>0$), the system yields at a (mechanical) yield strain $\gamma_{\rm y}^{\rm m}$, 
beyond which the stress $\sigma$ does not increase anymore.
At even larger strains, the system eventually evolves to a steady state where $\sigma$ reaches a constant, consistent with the experimental measurement in  Fig.~\ref{fig:experiment}C.


\subsubsection{Frictionless spheres: rapidly quenched}
\label{sec:rapid_quench}
In granular matter, the inter-particle friction is in general non-negligible.
The friction can create complex effects such as 
history-dependent dynamics, which challenges theoretical investigations. 
To understand whether the origin  of shear jamming is essentially associated to friction, 
many studies have focused on a simpler model: athermal frictionless spheres interacting via soft (harmonic or Hertzian) potentials.
The onset of jamming in athermal soft spheres by either compression or shearing is indicated by the emergence of rigidity (non-zero pressure, stress and moduli), following certain critical scaling laws (see Sec.~\ref{sec:scaling})

It was believed that shear jamming could not occur without inter-particle friction: more precisely, according to finite-size analyses of frictionless simulation data,  compression  and shear jamming have to occur at the same density $\varphi_{\rm J}$ in the thermodynamic limit~\cite{bertrand2016protocol, baity2017emergent}. In other words, it is impossible to jam frictionless spheres by shear deformations without compression. 
If this picture were valid, then $\varphi_{\rm J}$ is a unique boundary between unjammed and jammed configurations; imposing anisotropy to the configuration does not  help stabilizing the system. Such a scenario seems consistent with the absence of dilatancy observed in simulated frictionless packings~\cite{peyneau2008frictionless, azema2015internal}. However, in all these simulations~\cite{peyneau2008frictionless, azema2015internal, bertrand2016protocol, baity2017emergent}, the initial unstrained states are quenched rapidly from random configurations. While such rapidly quenched  systems have a unique jamming density $\varphi_{\rm J}$, generally the jamming density $\varphi_{\rm j}$ depends on protocol parameters such as the quench rate~\cite{speedy1996distribution, torquato2000random, chaudhuri2010jamming, ozawa2012jamming} (see Sec.~\ref{sec:J_plane}). For an annealed  system with a jamming density $\varphi_{\rm j}$ above $\varphi_{\rm J}$, the situation is essentially different, as we discuss below.  

\subsubsection{Frictionless spheres: deeply annealed}

\begin{figure}[!htbp]
  \centering
  \includegraphics[width=\linewidth]{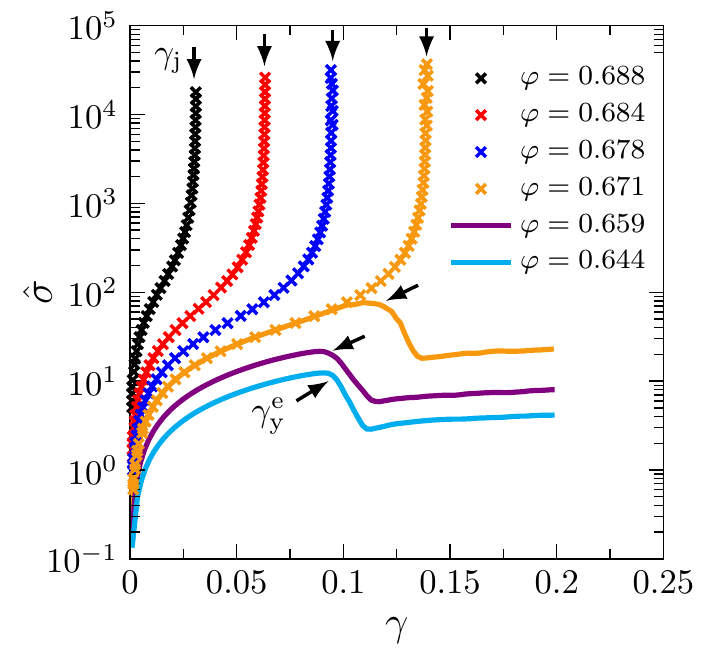}
  \caption{{\bf Shear jamming in simulations of   frictionless thermal hard spheres.} Reduced stress-strain curves 
  of  polydisperse hard spheres  at different densities, quenched from 
  $\varphi_{\rm eq} = 0.643$. 
  The density $\varphi_c \approx 0.671$ separates two types of behavior: for $\varphi \geq \varphi_{\rm c}$, the system jams at $\gamma_{\rm j}$ under shear (points); for $\varphi \leq \varphi_{\rm c}$, the system yields at $\gamma_{\rm y}^{\rm e}$ (lines). Near $\varphi_{\rm c}$, both types coexsit. The data are adapted from~\cite{jin2021jamming}.  
 }
  \label{fig:stress_strain_HS}
\end{figure}

\begin{figure*}[!htbp]
  \centering
  \includegraphics[width=0.65\linewidth]{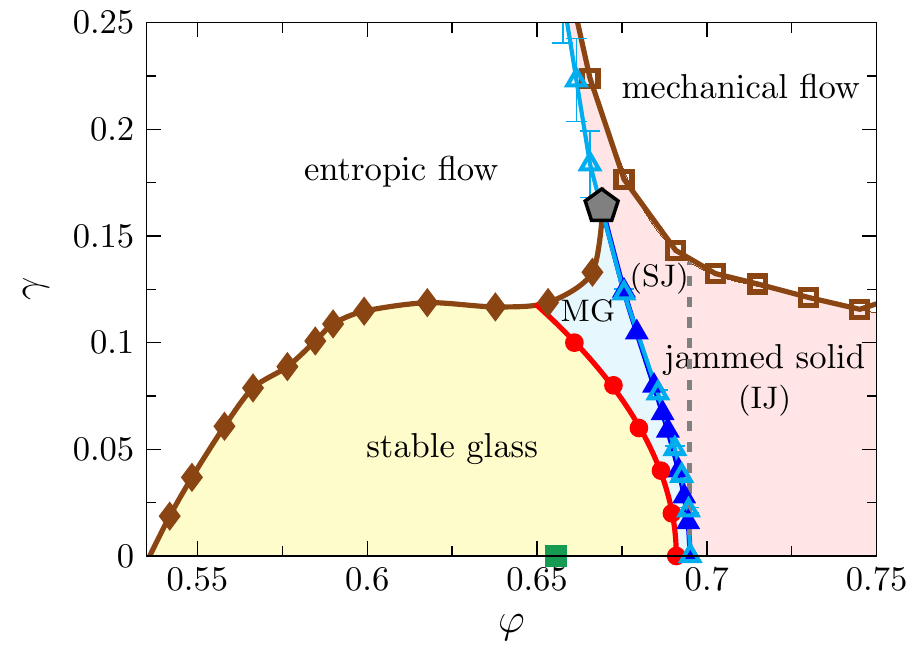}
  \caption{{\bf Unified strain-density phase diagram of shear jamming, yielding 
 and Gardner transitions in athermal soft spheres and thermal hard spheres.}
  Data are obtained for the polydisperse sphere model~\cite{jin2018stability}, with $\varphi_{\rm eq} = 0.655$ (green square), $\varphi_{\rm j} = 0.695$ (vertical dashed line), $\varphi_{\rm c} = 0.669$ (gray pentagon) and {$\varphi_{\rm J}\approx 0.655$}.
Open cyan triangles represent the shear jamming line $\gamma_{\rm j}(\varphi)$ obtained in athermal soft spheres, and  filled blue triangles represent the shear jamming line obtained in thermal hard spheres. 
Note that this is only an example of the shear jamming line with $\varphi_{\rm j} = 0.695$ in the J-plane (see Fig.~\ref{fig:Jplane} for other shear jamming lines).
On the left side of the shear jamming line is the entropic side for thermal hard spheres. 
The Gardner transition line $\gamma_{\rm G}(\varphi)$ (red circles)
separates the marginally stable   glass (MG) phase  from the stable glass phase.
Beyond the entropic yielding line $\gamma_{\rm y}^{\rm e}(\varphi)$ (brown diamonds) is the entropic flow phase. 
On the right side of the shear jamming line is the mechanical side for athermal soft spheres. The mechanical yielding line $\gamma_{\rm y}^{\rm m}(\varphi)$ (brown squares) separates the jammed solid phase and the mechanical flow phase. Above $\varphi_{\rm j}$, the system is isotropically jammed (IJ) without shear ($\gamma =0$), while for $\varphi_{\rm J} <  \varphi < \varphi_{\rm j}$ it is shear jammed (SJ) at $\gamma_{\rm j}$.
}
  \label{fig:stability_map}
\end{figure*}

To observe shear jamming in frictionless spheres, firstly one needs to prepare a  deeply annealed state with a jamming density $\varphi_{\rm j}> \varphi_{\rm J}$.
To realize that, we employ the swap algorithm, which has been widely used in Monte-Carlo simulations of particle systems~\cite{kranendonk1991computer, grigera2001fast, gutierrez2015static, berthier2016equilibrium, berthier2016growing, ninarello2017models}, on a frictionless polydisperse model in three dimensions~\cite{berthier2016equilibrium, berthier2016growing, ninarello2017models}. The swap algorithm generates equilibrium liquid configurations of thermal hard spheres at a large density $\varphi_{\rm eq}$ above the estimated dynamical glass transition density $\varphi_{\rm d}$ (also called mode-coupling theory (MCT) density, $\varphi_{\rm MCT} = \varphi_{\rm d}$). Compressing the system then results in jammed configurations with a large jamming density,  $\varphi_{\rm j}(\varphi_{\rm eq}) > \varphi_{\rm J}$ (see Fig.~\ref{fig:J_line}A). Simulation results obtained for such deeply-annealed frictionless 3D systems are presented in Figs.~\ref{fig:J_line}A, \ref{fig:Jplane}, \ref{fig:frictional_simulation}B, \ref{fig:stress_strain_HS}, \ref{fig:stability_map}, \ref{fig:two_step_yielding}, \ref{fig:schematic_PD} and~\ref{fig:elasticity_theory}. Details on the model and simulation methods can be found in Refs.~\cite{jin2018stability, jin2021jamming, babu2021dilatancy, pan2022nonlinear, pan2022shear}.

Once the initial jammed configurations are prepared, one can switch to soft  potentials (here we use the repulsive harmonic potential) and set a zero temperature ($T=0$) to simulate athermal shear of soft spheres.
Decompressing the system to a density $\varphi$ between $\varphi_{\rm J}$ and $\varphi_{\rm j}$  generates an unjammed state without a shear strain ($\gamma = 0$). Increasing $\gamma$ at the fixed $\varphi$ jams the system at the jamming strain $\gamma_{\rm j}$, where the stress $\sigma$ becomes non-zero (see Fig.~\ref{fig:frictional_simulation}B). 
In fact, the frictional (Fig.~\ref{fig:frictional_simulation}A) and frictionless (Fig.~\ref{fig:frictional_simulation}B) stress-strain curves show very similar behavior. 
Finite-size analysis confirm that, for deeply annealed frictionless systems, shear and compression jamming can indeed occur at different densities even in the thermodynamic limit~\cite{jin2021jamming}, in sharp contrast to the case of rapidly quenched systems as discussed in Sec.~\ref{sec:rapid_quench}.
These results demonstrate that the inter-particle friction is not  essential for shear jamming~\cite{ kumar2016memory, urbani2017shear,  jin2021jamming, babu2021dilatancy}. In Sec.~\ref{sec:phase_diagram}, we show that the origin of shear jamming in frictional and frictionless systems can be  explained universally by a common mechanism, which results in a generalized jamming phase diagram at zero temperature~\cite{babu2021dilatancy}. 



\subsection{Thermal hard spheres}

\label{sec:frictionless_HS}




For thermal hard spheres (without friction), the mean-field theory  predicts the possibility of shear jamming in deeply annealed ($\varphi_{\rm j} > \varphi_{\rm J}$) systems~\cite{urbani2017shear}. 
The compression jamming transition occurs at the jamming density $\varphi_{\rm j}$, 
accompanied by a divergence of the reduced pressure $\hat{P}$. 
If decompressed to a density $\varphi < \varphi_{\rm j}$, the system is  unjammed without strain deformations ($\gamma = 0$). 
Figure~\ref{fig:stress_strain_HS} reports reduced stress-strain ($\hat{\sigma}$-$\gamma$) curves of a simulated hard sphere model in three dimensions, with a very small shear rate (quasi-static shear)~\cite{jin2021jamming}. 
For $\varphi_{\rm c} < \varphi < \varphi_{\rm j}$, shear jamming occurs at a jamming strain $\gamma_{\rm j}$ where both reduced pressure and stress diverge.  For $\varphi < \varphi_{\rm c}$, shear yielding
takes place at an (entropic) yield strain $\gamma_{\rm y}^{\rm e}$. The two kinds of  behavior, shear yielding and shear jamming, are separated by a density $\varphi_{\rm c}$. These observations are qualitatively consistent with the predictions given by the mean-field theory~\cite{urbani2017shear}. 




\section{Connection between rheology of athermal soft spheres and thermal hard spheres via shear jamming}
\label{sec:connection_athermal_thermal}


From the discussion in Sec.~\ref{sec:simulation}, one finds that the rheological behavior of athermal soft spheres and  thermal hard spheres can be very different, due to the distinct nature of inter-particle interactions.
 In athermal soft spheres, the inter-particle force is non-zero only when the two particles are {\it mechanically} in contact, and thus the stress $\sigma$ and  pressure $P$ are strictly zero in the unjammed phase (see Fig.~\ref{fig:frictional_simulation}B; here we consider quasi-static shear of frictionless particles,  ignoring transient and frictional effects).  In thermal hard spheres, the effective forces between particles are due to collisions, and thus  the reduced  stress $\hat{\sigma}$ and pressure $\hat{P}$ come from pure {\it entropic} effects. Approaching shear jamming, the divergence of $\hat{\sigma}$ and  $\hat{P}$ (see Fig.~\ref{fig:stress_strain_HS}) implies that,  the gaps between nearest hard sphere neighbours vanish and the effective forces diverge due to infinitely frequent collisions.

The rheology of the two systems has been investigated separately in most existing studies. In typical experiments of granular matter~\cite{bi2011jamming, zhao2019shear, zhao2022ultrastable} (see Sec.~\ref{sec:granular_matter}) and simulations of athermal models~\cite{kawasaki2020shear, babu2021dilatancy, pan2022nonlinear} (see Sec.~\ref{sec:frictionless_SS}), the particles are at rest before jamming, and thus the physics related to  entropic effects cannot be explored. On the other hand, hard spheres cannot overlap, which means that the over-jammed regime is inaccessible in theories~\cite{urbani2017shear, altieri2019mean} (see Sec.~\ref{sec:granular_matter}) and simulations~\cite{jin2017exploring, jin2018stability} of thermal hard spheres  (see Sec.~\ref{sec:frictionless_HS}). Remarkably, connections can be made via shear jamming, providing a unified phase diagram Fig.~\ref{fig:stability_map} for the two systems,  which we discuss in detail below. Note that the entropic side of Fig.~\ref{fig:stability_map} is called a stability map for hard sphere glasses in Refs.~\cite{jin2018stability, altieri2019mean}, and the mechanical side corresponds to the generalized Liu-Nagel jamming phase diagram at zero temperature~\cite{babu2021dilatancy} as we discuss in Sec.~\ref{sec:phase_diagram}.

(i) Entropic side of the phase diagram for thermal hard spheres. We start from a deeply equilibrated liquid configuration of thermal hard spheres at $\varphi_{\rm eq}$. 
The system is isotropically jammed at a $\varphi_{\rm eq}$-dependent jamming density $\varphi_{\rm j}$ by compression.  
At a density $\varphi$ between $\varphi_{\rm eq}$ and $\varphi_{\rm j}$, the system is in an ultra-stable glass state: the motion of particles is {\it caged} by the neighbours,  and the entire system behaves like an amorphous solid with a finite shear modulus. Because the glass is deeply annealed (large $\varphi_{\rm eq}$), the structural relaxation  time  (the so-called $\alpha$-relaxation time~\cite{gotze2009complex}) is significantly large, and within that time scale, the solid is well-defined. If the glass were poorly annealed (small $\varphi_{\rm eq}$), then the rheological behavior such as yielding would be interfered with by the $\alpha$-relaxation process  --  here we do not consider such complicated situations. 

The ultra-stable hard sphere glass undergoes a {\it Gardner transition}~\cite{charbonneau2014fractal, berthier2019gardner, jin2022computer} with compression or shear. For $\gamma < \gamma_{\rm G}$, where $\gamma_{\rm G}$ is the Gardner transition strain, the
 system is in a {\it stable glass} state, and its
response  to the shear strain is elastic. For $\gamma > \gamma_{\rm G}$, the system becomes a {\it marginally stable glass}, indicated by the onset of plastic response to small strain increments~\cite{jin2018stability} and a protocol-dependent shear modulus~\cite{nakayama2016protocol, jin2017exploring, yoshino2014shear}. The above behavior essentially reflects a change of the free-energy landscape: a meta-stable glass basin splits into hierarchical sub-basins  in the marginal glass state. From a theoretical point of view, the Gardner transition belongs to the full-replica symmetry breaking universality class, initially discovered by Parisi in spin glass models~\cite{mezard1987spin}. The interested reader may refer to recent reviews~\cite{berthier2019gardner, jin2022computer} and references therein for further details on the Gardner transition.

As shown already  in Fig.~\ref{fig:stress_strain_HS}, upon increasing $\gamma$, the hard sphere glass either jams (for $\varphi > \varphi_{\rm c}$) or yields (for $\varphi < \varphi_{\rm c}$). The shear jamming of hard spheres at $\gamma_{\rm j}$  always occurs in the marginally stable phase. However, the system remains in the same free-energy glass basin (but not in the same sub-basin!). This kind of jamming is called {\it reversible jamming}~\cite{jin2021jamming}; it is reversible in the sense that the system remains in the same basin upon reverse shear. 
Once yielding occurs, the system escapes the glass basin.  After yielding ($\gamma > \gamma_{\rm y}^{\rm e}$), the system enters an {\it entropic flow} state, where the sample-averaged $\hat{\sigma} \sim \sigma/T$ is a positive constant, implying continuous cage breaking and reformation. We emphasize that the entropic flow is unjammed, since particles are not in permanent mechanical contact and $\sigma = 0$ if the temperature is turned off.

\begin{figure*}[!htbp]
  \centering
  \includegraphics[width=0.9\linewidth]{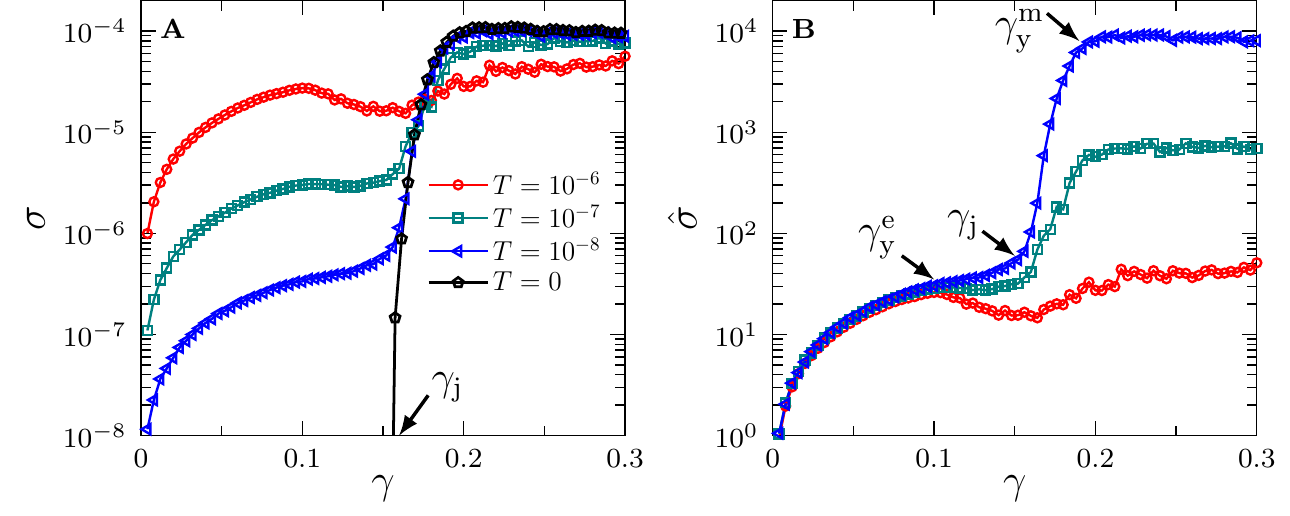}
  \caption{{\bf Shear jamming and two-step yielding.} 
  Data are obtained for the polydisperse sphere model~\cite{jin2018stability, jin2021jamming}, with { $\varphi_{\rm eq} = 0.655$, $\varphi_{\rm j} = 0.695$ and $\varphi_{\rm c}=0.669$.}
  Athermal quasi-static shear is applied for soft spheres with a constant $\varphi = 0.665$ below $\varphi_{\rm c}$.
 (A)  The stress $\sigma$  and (B) the  reduced stress $\hat{\sigma} = \sigma V/Nk_{\rm B}T$ are measured as functions of $\gamma$ for the configurations obtained by the athermal quasi-static shear, with a given temperature $T$ switched on.
The entropic yielding, shear jamming, and mechanical yielding occur successively at $\gamma_{\rm  y}^{\rm e}$, $\gamma_{\rm  j}$ and $\gamma_{\rm  y}^{\rm m}$, as indicated. 
  }
  \label{fig:two_step_yielding}
\end{figure*}

(ii) Mechanical  side of the phase diagram for athermal soft spheres. The initial states are prepared in the same way as described in (i) for thermal hard spheres. Once a hard sphere ultrastable glass  is generated, the temperature is turned off, and the soft  potential is switched on.  Shear jamming takes place at $\gamma_{\rm j}$ when the stress $\sigma$ becomes non-zero under quasi-static shear (see Fig.~\ref{fig:frictional_simulation}), where a stable contact network is formed. Yielding occurs at $\gamma_{\rm y}^{\rm m}$, beyond which the system enters a {\it mechanical flow} state. The positive constant (sample-averaged) stress $\sigma$ of the mechanical flow suggests  continuous 
breaking and reformation of the force network.

For $\varphi > \varphi_{\rm c}$, thermal hard spheres and athermal soft spheres jam at the same $\gamma_{\rm j}$, while for $\varphi_{\rm J} <  \varphi < \varphi_{\rm c}$, only athermal soft spheres can be shear jammed at a finite $\gamma_{\rm j}$. The shear jamming in this regime ($\varphi_{\rm J} <  \varphi < \varphi_{\rm c}$) requires the system to leave the original free-energy glass basin, which cannot be returned to by reverse shear (called {\it irreversible jamming}~\cite{jin2021jamming}). We note that the mechanical side of Fig.~\ref{fig:stability_map} looks similar  to the phase diagram of the reversible-irreversible transitions in athermal soft spheres~\cite{nagasawa2019classification}.


Interestingly, it is possible to connect the above two sides in a single procedure, as shown in Fig.~\ref{fig:two_step_yielding}. We apply quasi-static shear to athermal soft spheres, and save the configurations at each strain step. 
Then the stress $\sigma$ of the saved configuration at $\gamma$ is measured with a given $T$. In this way, although the configurations are obtained by athermal quasi-static shear, in the evaluation  of $\sigma$ both entropic and mechanical contributions have been included. 
Note that the data for $T \to 0$ converge to those presented in Fig.~\ref{fig:frictional_simulation}B. For a small but finite $T$, the $\hat{\sigma}$-$\gamma$ curves reveal a peculiar process of ``yielding-jamming-yielding", which occur at $\gamma_{\rm y}^{\rm e}$, $\gamma_{\rm j}$ and  $\gamma_{\rm y}^{\rm m}$ successively. It demonstrates that, the unjammed configurations below $\gamma_{\rm j}$ are in fact in glass states, consistent with the simulations of thermal hard spheres (Sec.~\ref{sec:frictionless_HS}). 
We notice that the stress-strain ($\sigma-\gamma$) curves in Fig.~\ref{fig:two_step_yielding}(A) look similar to the yield stress-density ($\sigma_{\rm y}-\varphi$) curves  reported previously in Ref.~\cite{ikeda2012unified} (see Fig.~2(b) in~\cite{ikeda2012unified}).
In both cases, the curves are $T$-dependent below jamming ($\gamma < \gamma_{\rm j}$ in Fig.~\ref{fig:two_step_yielding}(A) and $\varphi < \varphi_{\rm J}$ in ~\cite{ikeda2012unified}), and coincide  above jamming for small $T$, suggesting that glass rheology  and jamming rheology  are different but related~\cite{ikeda2012unified, mason1997osmotic}. While Ref.~\cite{ikeda2012unified}  focuses on rapidly quenched systems ($\varphi_{\rm j} \approx \varphi_{\rm J}$), it will be interesting to extend 
the analysis to the case with $\varphi_{\rm j} > \varphi_{\rm J}$.

Previous experiments~\cite{pham2006yielding, koumakis2011two, laurati2011nonlinear} and theories~\cite{altieri2018microscopic}   have noticed  a similar two-step yielding process (although without shear jamming) in attractive colloids.
In  attractive colloids, two-step yielding is originated from two distinct length scales in the interaction potential: one for the hard-core repulsion and the other one for the attractive tail~\cite{sellitto2013thermodynamic, altieri2018microscopic}. Although the potential $V(\delta)$ of our model only contains one length scale $\delta$ (the linear overlap between particles), at a finite temperature below jamming, there is actually another ``hidden" length scale, which is the linear gap $h$ between particles. At the jamming transition, both length scales, $\delta$ and $h$, vanish coincidentally. Thus the mechanism of two-step yielding in our system (Fig.~\ref{fig:two_step_yielding}) is essentially similar to that in attractive colloids. 
It 
is of interest
to examine if shear jamming can be incorporated with two-step yielding in colloids with a properly designed interaction.

\section{Universal description of shear jamming in frictionless and frictional  systems in a generalized zero temperature Liu-Nagel jamming phase diagram}
\label{sec:phase_diagram}

\begin{figure*}[!htbp]
  \centering
 \includegraphics[width = 0.9\linewidth]{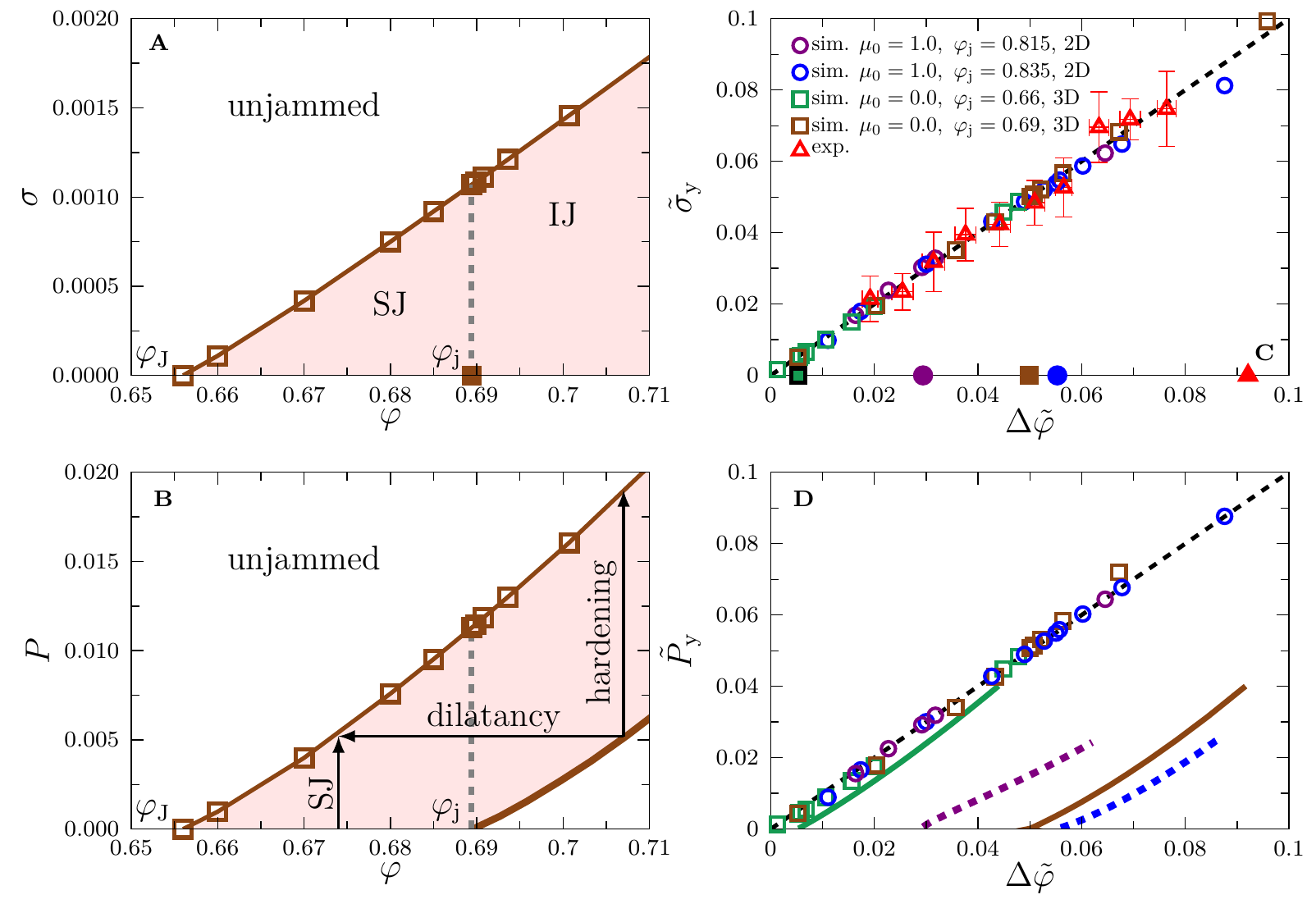}
  \caption{{\bf Generalized Liu-Nagel jamming phase diagram at zero temperature.}
  (A) Stress-density ($\sigma-\varphi$) and (B) pressure-density ($P-\varphi$) phase diagrams, obtained from simulations of the 3D frictionless polydisperse model, with $\varphi_{\rm j} = 0.69$ and $\varphi_{\rm J} \approx 0.655$~\cite{babu2021dilatancy, pan2022shear}.
  Open brown squares represent $\sigma_{\rm y}(\varphi)$ and $P_{\rm y}(\varphi)$ in (A) and (B) respectively. Isotropic jamming (IJ) and shear jamming (SJ) are separated at $\varphi_{\rm j}$ (vertical dashed line).
  In (B),  the thick solid brown line represents $P_0(\varphi)$ for the isotropically compressed equation of states; the black lines with arrows indicate typical routes for shear jamming, dilatancy and shear hardening respectively. 
  (C) Collapsing of the rescaled equation of states  $\Tilde{\sigma}_{\rm y} (\Delta\Tilde{\varphi})$ obtained from different systems (open symbols), 
  where $\Delta\Tilde{\varphi} =\varphi/\varphi_{\rm J} - 1$ and  $\Tilde{\sigma}_{\rm y} = \sigma_{\rm y}/C$.
  Here $\varphi_{\rm J}$ and $C$ depend on the system,  obtained from fitting the yield stress data to $\sigma_y(\varphi) = C (\varphi / \varphi_{\rm J} - 1)$.
  The filled symbols represent $\Delta\Tilde{\varphi}_{\rm j} = \varphi_{\rm j}/\varphi_{\rm J} -1$.
  The frictional simulations are performed for the 2D bidisperse model~\cite{pan2022nonlinear}.
  The experimental data of
  frictional granular disks
  are adapted from Ref.~\cite{bi2011jamming}.    
(D) Collapsing of $\Tilde{P}_{\rm y} (\Delta\Tilde{\varphi})$, where
  $\Tilde{P}_{\rm y} = P_{\rm y}/C'$ and  $C'$ is a fitting parameter in $P_{\rm y}(\varphi) = C'(\varphi/\varphi_{\rm J}-1)$.
  The lines (solid for 3D and dashed for 2D) represent the corresponding $P_0(\Delta\Tilde{\varphi})$, which is non-unique  in the rescaled plot. 
  }
  \label{fig:schematic_PD}
\end{figure*}


In this section, we focus on shear jamming at zero temperature ($T=0$) with and without inter-particle friction. Let us first 
discuss  athermal frictionless systems. 
In Fig.~\ref{fig:stability_map}, the mechanical yielding line $\gamma_{\rm y}^{\rm m}(\varphi)$ separates jammed solid states and unjammed mechanical flow states. In other words, the system is only jammed when its stress $\sigma$ is below the yield stress $\sigma_{\rm y}$ in a stress-controlled protocol; once $\sigma$ exceeds $\sigma_{\rm y}$, the system starts to flow. The equation of state, $\sigma_{\rm y}(\varphi)$,  
is the boundary between jammed and unjammed phases, in the  Liu-Nagel jamming phase diagram at zero temperature~\cite{liu1998nonlinear} (in a pressure-density phase diagram, the boundary is the yield pressure line $P_{\rm y}(\varphi)$). In this original proposal,  $\sigma_{\rm y}(\varphi)$ vanishes at the unique J-point, i.e., $\sigma_{\rm y} (\varphi_{\rm J}) = 0$.
The system is jammed by isotropic compression at a density $\varphi > \varphi_{\rm J}$, and is unjammed by shear upon yielding at $\sigma = \sigma_{\rm y}$ with $\varphi$ fixed above $\varphi_{\rm J}$. Unfortunately, such a picture cannot describe shear jamming. 

The Liu-Nagel phase diagram is later generalized to include shear jamming~\cite{bi2011jamming, babu2021dilatancy} (see Fig.~\ref{fig:schematic_PD}), which is essentially a stress-density (or pressure-density)  representation of the mechanical side of the strain-density phase diagram Fig.~\ref{fig:stability_map}.
Compared to the original proposal~\cite{liu1998nonlinear}, where the unique jamming density at $\varphi_{\rm J}$ is assumed, 
the generalized version takes into account the fact that the jamming density $\varphi_{\rm j}$  is non-unique (see Sec.~\ref{sec:J_plane}). 
For $\varphi > \varphi_{\rm j}$, the system is isotropically jammed due to compression,
while for $\varphi_{\rm J} < \varphi < \varphi_{\rm j}$, the system is unjammed without shearing and is  shear jammed at a finite $\gamma_{\rm j}$ (see Fig.~\ref{fig:stability_map}). The isotropically jammed states are protocol-dependent. For example,  their pressure-density relationship under compression, $P_0(\varphi) \sim \varphi-\varphi_{\rm j}$, depends on  $\varphi_{\rm j}$, i.e., how deep the initial configuration is annealed (here $P_0(\varphi)$ is linear near jamming because we consider the repulsive harmonic potential for simplicity, see Sec.~\ref{sec:scaling}).
However, the jamming-unjamming boundary is unique, i.e., $\sigma_{\rm y}(\varphi) \sim \varphi - \varphi_{\rm J}$ in the stress-density phase diagram, or $P_{\rm y}(\varphi) \sim \varphi - \varphi_{\rm J}$ in the pressure-density phase diagram, is independent of the initial condition ($\varphi_{\rm j}$), and 
of whether the system is compression or shear jammed. 
Indeed,  $\sigma_{\rm y}$ and $P_{\rm y}$ can  also be interpreted as the stress and pressure of steady states, where the  memory of the initial condition is completely lost. 
 Note that the lowest-density steady state at $\varphi_{\rm J}$ is known in soil mechanics as the {\it critical state}~\cite{wood1990soil}.

Frictionless and frictional spheres have significantly distinct properties at the microscopic level.  
The frictional forces can create torques that modify the mechanical equilibrium conditions. In addition, the frictional forces are in general history-dependent, unlike in frictionless spheres where the contact forces can be uniquely determined by the packing geometry at the jamming transition~\cite{charbonneau2015jamming}. 
Previous studies suggest that, friction may facilitate the formation and stabilization  of additional jammed states in sheared systems compared with the frictionless counterparts, which implies the essentialness of friction for shear jamming~\cite{vinutha2016disentangling}. However, as shown in Sec.~\ref{sec:simulation},
shear jamming could take place in both frictional and frictionless systems, encouraging us to seek a universal description. Indeed, independent of dynamical and mechanical properties at the particle level, the generalized Liu-Nagel jamming phase diagram also applies to frictional systems. With friction, the jamming-unjamming boundary  $\sigma_{\rm y} (\varphi)$ vanishes at a  density $\varphi_{\rm J}(\mu_0)$, which is presumably the lowest jamming density for the given friction coefficient $\mu_0$. Isotropically jammed states with a jamming density $\varphi_{\rm j} $ higher than  $\varphi_{\rm J}(\mu_0)$ can be created by protocols such as cyclic shear and tapping. However, once $\mu_0$ is fixed,  the boundary  $\sigma_{\rm y} (\varphi)$ or $P_{\rm y} (\varphi)$, is independent of $\varphi_{\rm j}$, just as in the frictionless case.
Applying the same argument used for frictionless systems above, 
one concludes that shear jamming would occur in the density regime $\varphi_{\rm J}(\mu) < \varphi < \varphi_{\rm j}$.


The universal description of shear jamming is further demonstrated by the rescaled  stress-density  plot in Fig.~\ref{fig:schematic_PD}C. 
The data of $\varphi_{\rm y}(\varphi)$, collected from 
experiments of frictional granular disks~\cite{bi2011jamming}, and athermal simulations in two and three dimensions, with and without friction, all collapse to a unique curve in the rescaled plot. Similar collapsing is found in the rescaled pressure-density plot (see Fig.~\ref{fig:schematic_PD}D). From these analyses, we find a general condition for shear jamming with ($\mu_0>0$) or without ($\mu_0=0$)  friction: one first locates the minimum jamming density $\varphi_{\rm J}(\mu_0)$ where $\sigma_{\rm y}$ vanishes, and then prepares configurations with a jamming density $\varphi_{\rm j} > \varphi_{\rm J}(\mu_0)$ using thermal annealing or mechanical training; for such systems, shear jamming shall be observed for $\varphi_{\rm J}(\mu_0) < \varphi < \varphi_{\rm j}$. Because shear brings a state with a large $\varphi_{\rm j}$ back to the generic state that jams at $\varphi_{\rm J}$ (see Fig.~4(A) in Ref.~\cite{pan2022nonlinear}), it effectively plays the opposite role of thermal annealing or mechanical training. 
Shear jamming is essentially related to this ``memory-erasing" procedure, which is independent of friction.

\section{Isostaticity and critical scalings of the shear jamming transition}
\label{sec:scaling}

In this section, we discuss isostaticity and criticality   near the shear jamming transition, focusing on frictionless systems. When the friction is present, the isostatic condition has to be modified~\cite{song2008phase, wang2011jamming}, and the critical exponents may become non-universal~\cite{wang2022experimental} -- these effects are less well understood and will not be discussed here. 
The critical scalings of the compression jamming transition have been intensively studied~\cite{makse2000packing, o2003jamming, zhang2005jamming, van2009jamming, liu2010jamming, wyart2005rigidity, charbonneau2014fractal, parisi2020theory}. 
Many of these scalings are preserved in shear jamming, but there are also exceptions. 


\subsection{Isostaticity}
\label{sec:isostaticity}
As in isotropic jamming, the shear jamming transition  always occurs under the isostatic condition~\cite{jin2021jamming}:
\beq
Z_{\rm j} =  2 d,
\label{eq:isostatic}
\eeq
where $Z_{\rm j}$ is the coordination number
in $d$ dimensions. In practice, rattlers, i.e., particles with fewer than $d+1$ contacts, should be removed from the count, and a finite-size correction should be considered~\cite{goodrich2012finite, goodrich2014jamming}.  Equation~(\ref{eq:isostatic}) can be derived by the Maxwell counting argument for minimum rigidity, which requires equality between 
the number of inter-particle forces and the number of force-balance equations~\cite{maxwell1870reciprocal, alexander1998amorphous}. This delicate balance between the degrees of freedom  and the number of constraints is the mechanical origin of criticality and marginality near jamming. For example,  the removal of a single bond from an isostatic system could result in a  floppy mode that 
extends in the entire system~\cite{wyart2012marginal}.
Thus approaching the jamming transition, there is a
 diverging mechanical length~\cite{silbert2005vibrations, wyart2005geometric}, which resembles the diverging  correlation length around a standard thermal critical point.

\subsection{Critical scalings: approaching shear jamming from below}

When the shear jamming transition is approached from below ($\gamma < \gamma_{\rm j}$) in thermal hard spheres, both the reduced pressure $\hat{P}$ and the reduced stress $\hat{\sigma}$ follow the free-volume scaling law~\cite{jin2018stability}:
\beq
\hat{P} \sim \hat{\sigma} \sim (\gamma_{\rm j} - \gamma)^{-1}.
\label{eq:scaling_hatP}
\eeq
The counterpart of Eq.~(\ref{eq:scaling_hatP}) in isotropic jamming is well-known~\cite{salsburg1962equation, donev2005pair, charbonneau2011glass}:
$\hat{P} \sim (\varphi_{\rm j} - \varphi)^{-1}$. The divergence of $\hat{P}$ and $\hat{\sigma}$  is a consequence of vanishing available (free) space for the particle motion, which suggests that the size of the particle's vibrational cage is reduced by shear. Generally, a planer shear deformation can be decomposed into two modes: compression in one direction and stretch in the perpendicular direction. The compression mode effectively confines the particle motion near shear jamming.

\subsection{Critical scalings: approaching shear jamming from above}

Before discussing the scalings above shear jamming ($\gamma > \gamma_{\rm j}$) in athermal soft spheres, it is useful to first summarize their counterparts in isotropic jamming, 
which have been accurately measured in many  studies~\cite{makse2000packing, o2003jamming, zhang2005jamming, van2009jamming, liu2010jamming, wyart2005rigidity}. Without loss of generality, we focus on the repulsive harmonic interaction  (i.e., the inter-particle potential is $V(\delta) \sim \delta ^2$ for contacting particles with a linear overlap $\delta$). The following scalings have been established when $\varphi$ approaches the isotropic jamming density $\varphi_{\rm j}$ from above. 

(I1)  A linear scaling between the pressure $P$ and the excess density $\Delta \varphi = \varphi - \varphi_{\rm j}$:
\beq
P \sim \Delta \varphi.
\label{eq:p_comp}
\eeq

(I2) A discontinuous change of the bulk modulus $B$ at the jamming transition: 
\beq
B \sim P^0.
\label{eq:B_comp}
\eeq

(I3) A square-root scaling of the excess coordination number $\Delta Z  = Z - Z_{\rm j} = Z-2d$: 
\beq
\Delta Z \sim \Delta \varphi^{1/2} \sim P^{1/2}.
\label{eq:Z_comp}
\eeq

(I4) 
The shear modulus $G$ vanishes continuously at the jamming transition: 
\beq
G \sim \Delta Z \sim P^{1/2}.
\label{eq:G_comp}
\eeq
 
The first two scalings Eqs.~(\ref{eq:p_comp}) and (\ref{eq:B_comp}) can be derived from a simple assumption of affine deformations (i.e., Hooke's law), in which case the pressure is proportional to the inter-particle force, $P \sim f$, and the bulk modulus is proportional to the stiffness $k$ (spring constant) of interaction bonds,   $B \sim k$. From the potential $V(\delta) \sim \delta ^2$, it is easy to obtain  that $f \sim \delta \sim \Delta \varphi$ and $k$ is a constant.

The affine arguments fail to explain the remaining two scalings, Eqs.~(\ref{eq:Z_comp}) and~(\ref{eq:G_comp}), since $\Delta Z = 0$ for pure affine deformations. They reflect an essential difference in the elastic response between  amorphous solids, where non-affine effects are inevitable for arbitrarily small deformations, and crystals, where the non-affinity can be safely neglected for small deformations. Equation~(\ref{eq:Z_comp}) has been derived from a scaling argument by imposing the condition of marginal stability in over-jammed systems, which also gives a diverging length scale at jamming $l \sim \Delta Z ^{-1}$~\cite{wyart2005rigidity}. 
The mean-field replica calculations with a full-step replica symmetry breaking ansatz also correctly reproduce the scaling Eq.~(\ref{eq:Z_comp})~\cite{charbonneau2014exact}.
The scaling Eq.~(\ref{eq:G_comp}), 
$G \sim \Delta Z$, can be understood from  the 
elasticity   theory of amorphous solids with non-affine effects taken into account~\cite{ wyart2005rigidity, lemaitre2006sum, zaccone2011approximate, zaccone2011network, karmakar2010athermal, yoshino2014shear, pan2022shear} (see Sec.~\ref{sec:elasticity}). 


For a general pow-law potential, $V(\delta) \sim \delta ^\alpha$, Eqs.~(\ref{eq:p_comp}-\ref{eq:G_comp}) become, 
(I1) $P \sim f \sim  
\Delta \varphi ^{\alpha-1}$, (I2) $B \sim k \sim \Delta \varphi^{\alpha -2}$, (I3) $\Delta Z \sim \Delta \varphi^{1/2}$, and (I4) $G \sim B \Delta Z \sim \Delta \varphi^{\alpha - 3/2}$. In particular, the universal exponent $1/2$ in the third scaling is independent of the interaction form, since it is originated from non-affinity. 

Next, let us consider the case of shear jamming. The control parameter of shear jamming can be either $\sigma$ or $\Delta \gamma = \gamma - \gamma_{\rm j}$, which quantifies the distance to the transition (analogously, in isotropic jamming the control parameter can be $P$ or $\Delta \varphi$). 
Corresponding to Eqs.~(\ref{eq:p_comp}-\ref{eq:G_comp}), the following scalings have been established~\cite{goodrich2016scaling, baity2017emergent, pan2022shear}.

(S1) A linear relationship between  stress and  strain,
\beq
\sigma \sim \Delta \gamma.
\label{eq:G_sj}
\eeq
This is nothing but the constitutive equation of elasticity, $\sigma = G \Delta \gamma$,  where  $G$ is the shear modulus.

(S2) A discontinuous change of the bulk modulus $B$ at the shear jamming transition,
\beq
B \sim \sigma^{0}.
\label{eq:B_sj}
\eeq

(S3) A square-root scaling of $\Delta Z$,
\beq
\Delta Z \sim \Delta \gamma ^{1/2} \sim \sigma ^{1/2}.
\label{eq:Z_sj}
\eeq

(S4) A discontinuous change of the shear modulus $G$ at the shear jamming transition,
\beq
G \sim \sigma^0.
\label{eq:G_sj}
\eeq

(S5) In addition, the macroscopic friction coefficient $\mu = \sigma /P$ (note that the microscopic inter-particle friction $\mu_0$ is always zero here) changes discontinuously at shear jamming due to the anisotropy of contact network at a non-zero $\gamma_{\rm j}$,
\beq
\mu \sim \sigma^{0}.
\label{eq:mu_sj}
\eeq
This is equivalent to  $P \sim \sigma$.


For the general power-law potential $V(\delta) \sim \delta ^\alpha$, the above scalings become,
(S1) $\sigma \sim f \sim \Delta \gamma ^{\alpha -1}$, (S2) $B \sim k \sim \Delta \gamma^{\alpha -2}$, (S3) $\Delta Z \sim \Delta \gamma^{1/2}$,  (S4) $G \sim k \sim \Delta \gamma ^{\alpha - 2}$, and (S5) $\mu \sim \Delta \gamma^0 \sim \sigma^0$. Again, the universal exponent $1/2$ in the third scaling remains unchanged. 




According to Eq.~(\ref{eq:G_sj}), the shear modulus $G$ changes discontinuously at shear jamming, 
in sharp contrast to the case of isotropic jamming, Eq.~(\ref{eq:G_comp}), where $G$ vanishes continuously at jamming. 
This difference is explained by the elasticity theory in Sec.~\ref{sec:elasticity}.
The last scaling Eq.~(\ref{eq:mu_sj}) is a consequence of the anisotropy of bond orientations denoted by a unit vector $\hat{n}(r)$ at a distance $r$. Taking a simple shear in the $x$-$z$ plane as an example, at shear jamming the 
 ``disconnected" correlation function $C(r) \sim \langle n^x(0) n^z(0) n^x(r) n^z(r) \rangle $  converges to a positive constant in the limit of large distance, $C(r \to \infty)  > 0 $. According to the {\it pressure–shear stress exponent equality}~\cite{goodrich2016scaling, baity2017emergent}, this gives $\sigma^2 \sim P^2$, or Eq.~(\ref{eq:mu_sj}). In isotropic jamming, $C(r \to \infty) = 0$ instead, which results in a different relationship, $\sigma^2 \sim P^2 /N$, where $N$ is the total number of particles in the system.  With Eqs.~(\ref{eq:Z_comp}) and~(\ref{eq:mu_sj}), the remaining scaling Eq.~(\ref{eq:Z_sj}) can be derived.

The shear jamming scalings discussed here apply to the  case when the jamming strain $\gamma_{\rm j}$ is non-zero~\cite{baity2017emergent}. When $\gamma_{\rm j} \to 0$, the scaling behavior appears to be more complex~\cite{pan2022shear}, which needs further exploration  and will not be discussed here. 

\subsection{Critical scalings: at the shear jamming transition}
At the transition, shear jamming and isotropic jamming share  some  universal scalings. 
The pair correlation function $g(r)$ has a delta peak at $r = 1$ (we set unit particle diameter) due to the formation of contacts. 
Near this singularity, the distribution of small gaps between near-contacting particles diverges following a power law,
\beq
g(r) \sim (r-1)^{-\alpha}.
\label{eq:g_scaling}
\eeq
The distribution of weak inter-particle forces  also follows a power law, with two different exponents:
\beq
P(f) \sim f^{\theta_{\rm e}},
\eeq
for forces associated with extensive modes,
and
\beq
P(f) \sim f^{\theta_{\rm l}},
\label{eq:fl_scaling}
\eeq
for those with localized buckling modes. It has been argued that the above exponents are interrelated via, $\theta_{\rm e} = 1/\alpha -2$ and  $\theta_{\rm l} = 1 - 2\alpha$, by satisfying the condition of marginal mechanical stability~\cite{wyart2012marginal, lerner2013low, degiuli2014force, muller2015marginal}.

The mean-field full replica symmetry breaking calculations~\cite{charbonneau2014fractal, charbonneau2014exact} give $\alpha = 0.41269$ and $\theta_{\rm e} = 0.42311$, which indeed satisfy the relationship $\theta_{\rm e} = 1/\alpha -2$.
If we take this value of $\alpha$  and combine it 
with the marginal  condition for localized buckling modes, $\theta_{\rm l} = 1 - 2\alpha$, then we obtain  $\theta_{\rm l} = 0.17462$. 
It has been found that these
exponents are consistent with 
finite-dimensional data in 
simulations~\cite{charbonneau2014fractal, charbonneau2015jamming} and experiments~\cite{wang2022experimental}  of compression jammed packings. 
Recent simulations report the same universal exponents for shear-jammed frictionless packings, in two and three dimensions~\cite{jin2021jamming, babu2022criticality}. However, an experiment of   sheared photoelastic  disks suggests a breakdown of this universality by inter-particle friction~\cite{wang2022experimental}.

\section{Elasticity of shear-jammed solids} 
\label{sec:elasticity}

\begin{figure*}[!htbp]
  \centering
 \includegraphics[width=\linewidth]{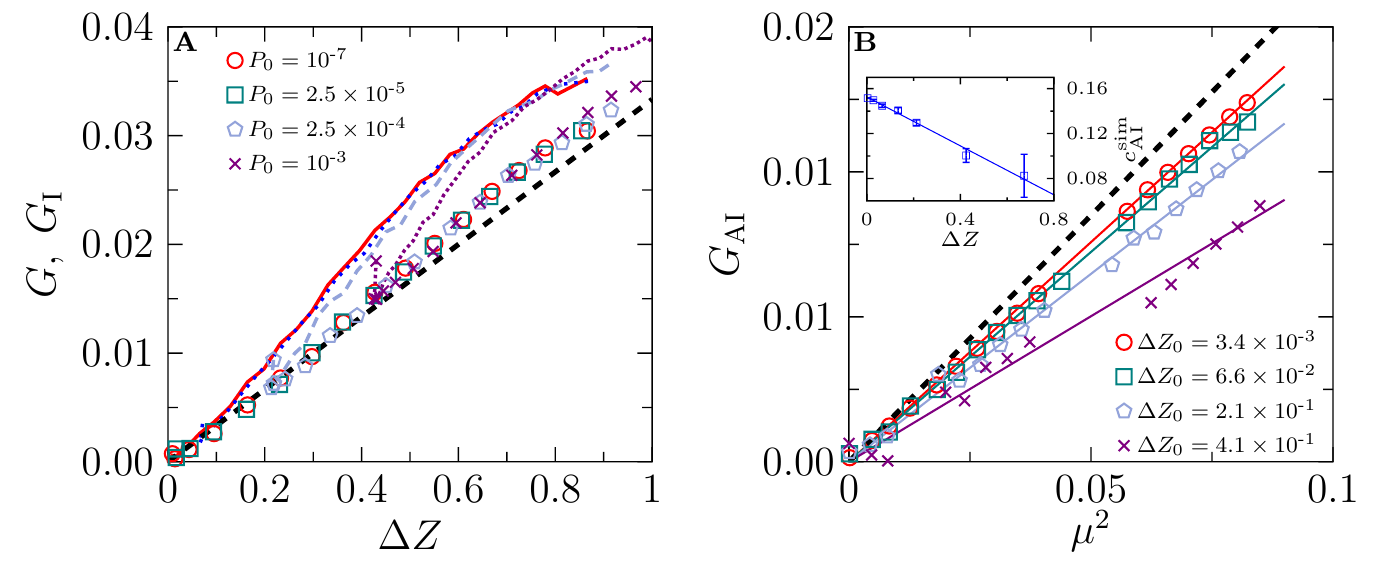}
  \caption{{\bf Shear modulus of amorphous solids near the shear jamming transition.}
  The data are obtained from athermal quasi-static shear  simulations of polydisperse soft spheres. The total shear modulus $G =G_{\rm I} + G_{\rm AI}$ at different shear strains is decomposed into the isotropic part $G_{\rm I}$ and the anisotropic part $G_{\rm AI}$.
  (A) Total shear modulus $G$ (lines) and its isotropic part $G_{\rm I}$ (points)
  as functions of $\Delta Z$. The pressure of unstrained initial states is indicated by $P_0$.
  The {bold} dashed line corresponds to the theoretical result $G_{\rm I} = c_{\rm I} \Delta Z$ with $c_{\rm I} =  1/30$.
  {(B) Anisotropic part of the shear modulus $G_{\rm AI}$  as a function of $\mu^2$ such that  $\Delta Z  = \Delta Z_0$ is fixed (note that $G_{\rm AI}$ is equal to $G-G_{\rm I}$ in (A)).
  The {bold} dashed line indicts the theoretical result $G_{\rm AI} = c_{\rm 0} \mu^2$ {for $\Delta Z \to 0$,} with   $c_{\rm 0}  \approx 0.17$ estimated from the force distribution.
  The solid lines represent the linear fitting, $G_{\rm AI}  = c_{\rm AI}^{\rm sim} \, \mu^2$, of the data. The slopes $c_{\rm AI}^{\rm sim}$ are  plotted in the inset,  which  can be fitted to a linear function  $c_{\rm AI}^{\rm sim}(\Delta Z) = 0.15 - 0.11 \Delta Z$ (line).} The data are adapted from Ref.~\cite{pan2022shear}.
  }
  \label{fig:elasticity_theory}
\end{figure*}

The jammed solids near the jamming transition ($P \to 0$ and $\Delta Z \to 0$) have rich mechanical properties due to their amorphous structures and marginality. 
The corresponding elasticity theory  has been developed in recent years~\cite{wyart2005rigidity, lemaitre2006sum, zaccone2011approximate, zaccone2011network, karmakar2010athermal, yoshino2014shear, pan2022shear}. For such a theory, one needs to extend the seminal work by Born and Huang that is  suitable only for solids   with pure affine responses (e.g., crystals)~\cite{born1954dynamical},  
properly taking into account non-affine effects.  Near the shear jamming transition, we should also consider  additional contributions from the  anisotropy~\cite{pan2022shear}, which is naturally neglected in isotropic jamming. Note that in this section, we 
focus on athermal and frictionless systems ($T=0$ and $\mu_0=0$).


According to the elasticity theory,  the bulk modulus $B$  depends linearly on the excess coordination number $\Delta Z$ that quantifies  the change of the number of contacts~\cite{pan2022shear},
\beq
B = c_0 + b\, \Delta Z,
\label{eq:B}
\eeq
where $b=1/18 + c_0/6$. The constant term $c_0 = \frac{\langle f \rangle^2}{3 \langle f^2 \rangle }$ is determined by the second normalized moment of the inter-particle force distribution $P(f)$, which is invariant under shear~\cite{pan2022shear}.
Note that this formula works for jammed solids near both isotropic and shear jamming, up to a first-order approximation. 
Because $c_0$ is non-zero, Eq.~(\ref{eq:B}) results in scalings Eq.~(\ref{eq:B_comp}) and~(\ref{eq:B_sj}), which reveal that 
 the onset of $B$ is discontinuous  at jamming when $\Delta Z \to 0$.

The shear modulus  $G$ depends on both $\Delta Z$ and the macroscopic friction coefficient $\mu = \sigma/P$ that quantifies the anisotropy~\cite{pan2022nonlinear},
\beq
G = G_{\rm I} + G_{\rm AI} = c_{\rm I} \, \Delta Z + c_{\rm AI} \, \mu^2,
\label{eq:G_theory}
\eeq
where $c_{\rm I} = 1/30$~\cite{zaccone2011approximate, zaccone2011network} and  $c_{\rm AI} = c_0 - a\, \Delta Z$ with a constant $a \approx 1$~\cite{pan2022nonlinear}. 
The expression Eq.~(\ref{eq:G_theory}) is well supported by simulation data as shown in Fig.~\ref{fig:elasticity_theory}.
According to Eq.~(\ref{eq:G_theory}), $G$  can be decomposed into two  parts: an isotropic part $G_{\rm I}  = c_{\rm I} \, \Delta Z$ and an anisotropic part $G_{\rm AI}  = c_{\rm AI} \, \mu^2$.
To the  lowest order, $G_{\rm I} = \frac{1}{30}\Delta Z$ and $G_{\rm AI} = c_0 \mu^2$, which  contribute independently to the total shear modulus.
The next  order correction is a crossover term, $-a\Delta Z \mu^2$, which gives a negative correction to $G$ in over-jammed packings ($\Delta Z > 0$).


Equation~(\ref{eq:G_theory}) uncovers the origin of the contradictory (continuous versus discontinuous-) behavior of the shear modulus $G$ near compression and shear jamming (see Eqs.~(\ref{eq:G_comp}) and~(\ref{eq:G_sj}))~\cite{baity2017emergent}. 
In compression jamming, since $\mu = 0$ (the system is always isotropic), only the isotropic term $G_{\rm I}$ is relevant~\cite{zaccone2011approximate, zaccone2011network}, and then Eq.~(\ref{eq:G_theory})  leads to the scaling Eq.~(\ref{eq:G_comp}) with a vanishing  $G$  at the transition.
 In contrast, because the shear jamming transition generally occurs at a non-zero strain $\gamma_{\rm j}$,
 the solid is anisotropic at the shear jamming transition, 
  i.e., $ \mu(\gamma_{\rm j}) > 0$ (see Eq.~(\ref{eq:mu_sj})). Then the anisotropic term $G_{\rm AI}$ in Eq.~(\ref{eq:G_theory})  gives a non-zero shear modulus $G = c_0 [\mu(\gamma_{\rm j}) ]^2$ at the transition even though $\Delta Z = 0$, resulting in the scaling Eq.~(\ref{eq:G_sj}).

\section{Related rheological  phenomena}
\label{sec:related}

\subsection{Connections of shear jamming  to dilatancy and shear hardening effects}
\label{sec:other}

The dilatancy effect, meaning that the volume of granular matter expands when subjected to shear deformations, was initially demonstrated by Reynolds over 100 years ago~\cite{reynolds1885dilatancy, rowe1962stress}. On the other hand, shear hardening (or strain hardening)  refers to the phenomenon of non-linear elasticity with the shear modulus increasing by strain~\cite{pan2022nonlinear, kawasaki2020shear, pan2022shear}.
The generalized Liu-Nagel jamming phase diagram in Fig.~\ref{fig:schematic_PD} provides a unified framework of shear jamming,   dilatancy and shear hardening in granular matter, revealing their deep connections. 

(i) Shear jamming. As described in Secs.~\ref{sec:connection_athermal_thermal} and~\ref{sec:phase_diagram}, shear jamming emerges along a constant-volume route at a fixed  $\varphi$, which satisfies $\varphi_{\rm J}  < \varphi < \varphi_{\rm j}$. Such a route starts from an unjammed configuration at $\gamma = 0$ with  $P=\sigma = \mu =\Delta Z = 0$, and ends in a steady state at a large strain ($\gamma \gg \gamma_{\rm y}$) with positive $P$, $\sigma$, $\mu$ and $\Delta Z$~\cite{babu2021dilatancy, pan2022nonlinear}.
Thus shear jamming has to occur at a certain $\gamma_{\rm j}$ during the route where these quantities become non-zero. 


(ii) Shear hardening. If one sets a constant-volume route at $\varphi > \varphi_{\rm j}$, then the initial state at $\gamma =0$ is already jammed --  the state variables $P$ and $\Delta Z$ are positive without shear.
As shown in 
Fig.~\ref{fig:schematic_PD}, in the steady state, $P$ and $\sigma$ are higher than 
their initial values,
which suggests that the solid is strengthened  by shear. In fact, similar to Fig.~\ref{fig:schematic_PD},  $\Delta Z$-$\varphi$ and $\mu$-$\varphi$ diagrams can be also constructed~\cite{babu2021dilatancy, pan2022shear}, showing that $\Delta Z$ and $\mu$ also increase along the route. Then according to the elasticity theory Eq.~(\ref{eq:G_theory}), the shear modulus $G$ should  increase with $\gamma$, resulting in shear hardening. 

(iii) Dilatancy. In constant-pressure shear protocols,
as can be seen from Fig.~\ref{fig:schematic_PD}C, the density $\varphi$ decreases, or in other words, the volume dilates. 

The above discussion demonstrates that shear jamming, shear hardening and dilatancy have a common origin (regardless of friction): shear 
destroys the optimized structure embedded in the initial packing, and the above three phenomena are all consequences (under different conditions) of this effect. 
If one regards the volume as an ``effective energy" of granular matter, as initially proposed by Edwards~\cite{baule2018edwards, edwards1989theory}, then  
shear increases the ``effective temperature", driving the system to a higher energy (lower jamming density $\varphi_{\rm j}$) state. 
It should be mentioned that other mechanisms may exist: e.g., a famous saw-tooth model has been proposed to explain  dilatancy in granular materials with friction~\cite{bolton1986strength}.
Also, 
interestingly cyclic shear seems to play the opposite role, since it increases the jamming density $\varphi_{\rm j}$~\cite{kumar2016memory, das2020unified, babu2021dilatancy}.

\subsection{Fragile states and their relation to discrete shear thickening}
\label{sec:fragile}


In experiments of dry granular matter, Bi et al. observed that  the force chains percolate throughout the shear-jammed system (e.g., see Fig.~\ref{fig:experiment}B), while prior to shear jamming, the system can enter a fragile state, where force chains only percolate in a single direction~\cite{bi2011jamming}.
 This fragile state exists between the unjammed and shear-jammed states, and closely resembles the theoretical model of fragile matter proposed by Cates et al.~\cite{cates1998jamming}. 
 In the model, fragile matter is represented by parallel straight force chains,
 which  can only bear compatible loads along the chain direction and fail catastrophically under any loads perpendicular to the chain direction. Zhao et al. 
 employed reverse shear to differentiate between fragile and shear-jammed states:  
 in fragile states, the shear stress $\sigma$ drops below the noise level  during reverse shear, whereas $\sigma$ remains above the noise level  in shear-jammed states~\cite{zhao2019shear} (this  definition does not distinguish between fragile and unjammed states as initially proposed in Ref.~\cite{bi2011jamming}).
The robustness of fragile states against particle rigidity (e.g., soft versus hard particles) and shear protocols (e.g., with a finite shear rate or temperature) 
remains an open question due to their delicate metastable characteristics.


Unlike dry granular materials, where shear jamming occurs at a zero strain rate, a sufficiently high shear rate is required in dense suspensions in order to observe shear jamming, 
because the particles have to overcome the lubrication force to get close to each other. 
The hydrodynamic force also sets the onset 
scale of shear stress in  discrete shear thickening.
In early works, it was conjectured that discrete shear thickening is  related to the hydro particle clusters, which were observed experimentally using confocal techniques~\cite{Cheng-2011imaging}. However, hydro clusters alone can only cause a mild stress increase;  they cannot account for the  sudden change of shear stress over several orders of magnitude during discrete shear thickening~\cite{Brown-Jaeger-Review}. 
For shear jamming and discrete shear thickening to occur, 
particles need to be brought close enough to each other under a sufficiently high strain rate, such that 
the inter-particle frictional force can provide mechanically stable support. 
Under suitable boundary conditions, when the anisotropic force chains percolate in one direction, discrete shear thickening occurs; when the force chains percolate through all boundaries, shear jamming occurs. 
In other words, discrete shear thickening corresponds to the fragile states discussed above. 
This physical picture is supported by experimental results \cite{peters2016direct}, and 
can explain both the rheological behavior of steady states
and transient phenomena, such as the impact of a heavy object onto the cornstarch suspension \cite{Waitukaitis-2012impact}. 


\section{Outlook}
\label{sec:outlook}



Since the first observation of shear jamming~\cite{zhang2010jamming, bi2011jamming}, it has been 
realized that the physics of shear jamming is extremely rich and deep. Many interesting questions remain to be answered. How would shear jamming depend on experimental protocols (e.g., uniform  versus boundary-driven shear)? What are the differences between shear jamming produced in a transient process and steady driving (e.g., continuous  versus cyclic shear)? 
What is the origin  of fragile states preceding shear jamming~\cite{bi2011jamming}? How would 
a finite shear rate affect shear jamming? What would be the influence of the non-spherical shape of particles?

It will be fascinating to perform three-dimensional experiments 
to study  shear jamming 
of dry granular materials in the microgravity environment~\cite{d2022granular}. So far,  direct experimental observations of shear jamming  have been only obtained in two-dimensional systems (granular disks). 
Three-dimensional experiments have been challenging 
because the gravity on earth naturally would cause an unsheared granular material to collapse onto a pile.
It is also of interest to understand 
what happens in dimensions above three.
Simulations can be carried out 
in four dimensions and above, which can then be compared to the theoretical results in large dimensions~\cite{urbani2017shear, altieri2019mean} to examine dimensional effects. 
Mean-field jamming models, such as the Mari-Kurchan model~\cite{mari2011dynamical}, could also be adapted to study shear jamming. 


On the theory side, there have been extensive  attempts to establish a statistical mechanics framework of jammed matter~\cite{edwards1989theory, makse2004statistical, bi2015statistical, parisi2020theory}.  The mean-field replica  approach, inherited from spin glass theories~\cite{parisi2010mean, charbonneau2017glass, parisi2020theory}, is based on presumed connections between liquid and jammed  ensembles.  
Experimental and simulation results on
shear jamming raise some 
fundamental challenges to the existing  theories.
The current state-following replica theory~\cite{rainone2016following, urbani2017shear, altieri2019mean} explains qualitatively the reversible shear jamming behavior, but does not capture irreversible shear jamming (Sec.~\ref{sec:connection_athermal_thermal}). 
The mechanism for the decrease  of jamming density $\varphi_{\rm j}$ (Sec.~\ref{sec:phase_diagram}) and the increase of configurational entropy (complexity) during shear jamming~\cite{pan2022nonlinear}  remains to be explained theoretically. 
The highly non-linear shear response around the onset of shear jamming  inspires one to investigate the associated fluctuations and  reexamine the non-equilibrium fluctuation-dissipation relations, which have been tested mainly in the flow regime in previous studies~\cite{makse2002testing, song2005experimental, zeng2022equivalence}.

Our ultimate goal is to understand the complex response of disordered systems to external fields. 
In this review, we focus on the assembly of spheres, which   represents a paradigm of disordered systems, and we consider the shear deformation, which  is a typical kind of external field that can be applied to such systems. Shear jamming is thus a phase transition introduced by the external field, instead of the temperature, in disordered systems. 
One may consider the famous de Almeida-Thouless transition~\cite{de1978stability}, a spin glass transition controlled by  an external magnetic field (at $T\to0$), as another example in this category. 
 Furthermore, the field can 
 also be self-generated.
Recent studies have revealed the similarities between 
global shear in granular matter and random local self-forcing in active matter~\cite{liao2018criticality, morse2021direct, agoritsas2021mean} -- 
thus shear jamming might be related to the dynamical phase transitions driven by
activity.


\begin{acknowledgments}
 We warmly thank Ada Altieri, Varghese Babu, Marco Baity-Jesi, Bulbul Chakraborty, Patrick Charbonneau,
 Sheng Chen, 
 Eric DeGiuli, Giampaolo Folena, Hisao Hayakawa, Takeshi Kawasaki,  Kunimasa Miyazaki, Corey S. O'Hern, Michio Otsuki, S. H. E. Rahbari, Srikanth Sastry, Ryohei Seto, Stephen Teitel, Pierfrancesco Urbani, Yujie Wang, Francesco Zamponi and Yiqiu Zhao for discussions.
  This work was supported 
  by the National Natural Science Foundation of China (Grants 12161141007, 11974361, 11935002,  12047503, 11974238 and 12274291), 
  the Chinese Academy of Sciences (Grants ZDBS-LY-7017 and KGFZD-145-22-13),
  the Wenzhou Institute (Grant WIUCASQD2023009),
  the Innovation Program of Shanghai Municipal Education Commission (No. 2021-01-07-00-02-E00138),
  and the KAKENHI (No. 25103005  ``Fluctuation \& Structure'' and  No. 19H01812) from MEXT, Japan.
\end{acknowledgments}

\bibliography{jamming}
\end{document}